%% file: staerk24c_dielectric-planar.tex
\documentclass[twocolumn,aps,superscriptaddress,longbibliography]{revtex4}
\usepackage{graphicx}
\usepackage[fleqn]{amsmath}
\usepackage{amssymb}
\usepackage[dvipsnames]{xcolor}
\usepackage[normalem]{ulem}
\definecolor{goodgreen}{rgb}{0.1,0.5,0}
\definecolor{goodred}{rgb}{0.7,0,0}
\usepackage{float}
\usepackage{multirow}
\usepackage{xr}
\usepackage{arydshln}

\usepackage{listings} 

\usepackage{array,mathtools,amssymb,booktabs}
\newcolumntype{C}{>{$}c<{$}}

\AtBeginDocument{%
  \heavyrulewidth=.08em
  \lightrulewidth=.05em
  \cmidrulewidth=.03em
  \belowrulesep=.65ex
  \belowbottomsep=0pt
  \aboverulesep=.4ex
  \abovetopsep=0pt
  \cmidrulesep=\doublerulesep
  \cmidrulekern=.5em
  \defaultaddspace=.5em
}

\usepackage[colorlinks,urlcolor=goodgreen,citecolor=blue,linkcolor=goodred]{hyperref}
\usepackage{bm}
\allowdisplaybreaks
\usepackage{todonotes} 
\usepackage{cleveref} 
\usepackage{siunitx} 
\usepackage{physics} 
\usepackage{csquotes} 

\usepackage{pgfplots} 

\usepackage{import}
\usepackage{xifthen}
\usepackage{pdfpages}
\usepackage{transparent}

\renewcommand{\vec}[1]{\bm{#1}}
\renewcommand{\epsilon}{\varepsilon}

\newif\ifshowhighlights

\showhighlightsfalse

\ifshowhighlights
    \newcommand{\hlork}[1]{\textcolor{red}{#1}}
\else
    \newcommand{\hlork}[1]{#1} 
\fi
\newcommand{\beginsupplement}{%
        \setcounter{table}{0}
        \renewcommand{\thetable}{S\arabic{table}}%
        \setcounter{figure}{0}
        \renewcommand{\thefigure}{S\arabic{figure}}%
        \setcounter{equation}{0}
        \renewcommand{\theequation}{S\arabic{equation}}
}

\begin{document}

\title{Static Dielectric Permittivity Profiles and Coarse-graining Approaches for Water in Graphene Slit Pores}
\author{Philipp Stärk}
\affiliation{Stuttgart Center for Simulation Science (SC SimTech),
  University of Stuttgart, 70569 Stuttgart, Germany}
\affiliation{Institute for Computational Physics, University of Stuttgart,
   70569 Stuttgart, Germany}
\author{Henrik Stooß}
\affiliation{
  Institute for Physics of Functional Materials,
  Hamburg University of Technology,
  21073 Hamburg, Germany
}
\author{Philip Loche}
\affiliation{Laboratory of Computational Science and Modeling, IMX, École
  Polytechnique Fédérale de Lausanne,1015 Lausanne, Switzerland}
\author{Douwe Jan Bonthuis}
\affiliation{Institute of Theoretical and Computational Physics,
  Graz University of Technology, 8010 Graz, Austria}
\author{Roland R.\ Netz}
\affiliation{Fachbereich Physik, Freie Universität Berlin,
	14195 Berlin, Germany}
\author{Alexander Schlaich}
\email{alexander.schlaich@tuhh.de}
\affiliation{
  Institute for Physics of Functional Materials,
  Hamburg University of Technology,
  21073 Hamburg, Germany
}

\date{\today}

\begin{abstract}
The dielectric response of nano-confined fluids is important for a wide range of technologies and biological systems, ranging from the fundamentals of energy storage to the stability of lipid membrane bilayers.
Its calculation from molecular simulations has become a viable tool since the
2010s and is often employed to rationalize experimental observations.
However, confusion has emerged on the underlying boundary conditions and
correspondingly the concept of dielectric profiles is often misunderstood in the literature.
Here, we re-derive the Green--Kubo relation for the linear dielectric response profile of fluids in planar confinement, while carefully considering the different underlying boundary conditions;
if the latter are correctly accounted for, profiles from equilibrium simulations are perfectly in line with results using externally applied fields.
We discuss common pitfalls and address misunderstandings found in
the literature, regarding both the detailed calculation and the general concept of dielectric profiles, as well as procedures to coarse-grain the microscopic dielectric behavior in order to connect to experimental observables.

Simulations consistently reveal that the dielectric response of water is bulk-like down to confining scales as small as $\sim 1$~nm. However, the effective dielectric response of confined systems, which determines their capacitance, depends on the precise location of the dielectric interface. Using concepts from effective medium theory, we show that the long-range reduction of the effective dielectric response, reported in both experimental and theoretical studies, is in line with a dielectric response that reaches the bulk value at about 1 nm from the surface. The effective dielectric response can be interpreted in terms of the commonly employed concept of an interfacial capacitance. We find that the dielectric properties of simulated water are independent of the simulation details and the water model employed, thus revealing the universal properties of water polarization correlations.

\end{abstract}

\maketitle

\input{document.tex}

\bibliography{MyLibrary2.bib,bib_as.bib}

\onecolumngrid
\clearpage

\renewcommand{\thefigure}{S\arabic{figure}}
\renewcommand{\theequation}{S\arabic{equation}}
\setcounter{figure}{0}
\setcounter{equation}{0}
\setcounter{section}{0}

\beginsupplement
\section{Supplemental Figures}

\begin{figure}[h]
    \centering
    \includegraphics[width=0.49\textwidth]{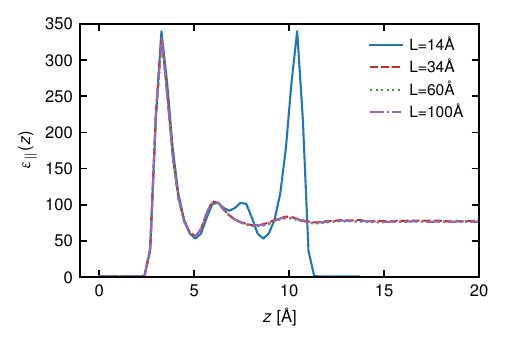}
    \includegraphics[width=0.49\textwidth]{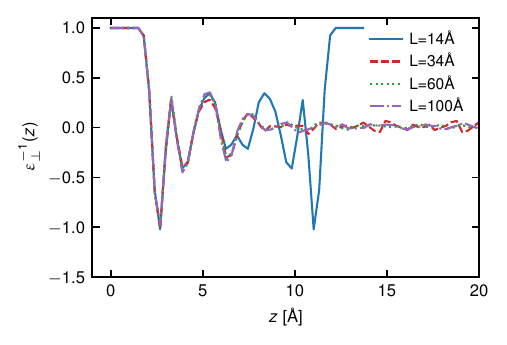}
    \caption{Comparison of the interfacial region of dielectric profiles of the TIP4P/$\epsilon$ water model. The different system sizes are plotted on top of each other to highlight how similar their profiles are near the interface.}
    \label{fig:box_size_comparison}
\end{figure}

\begin{figure}[h]
    \centering
    \includegraphics[width=0.49\textwidth]{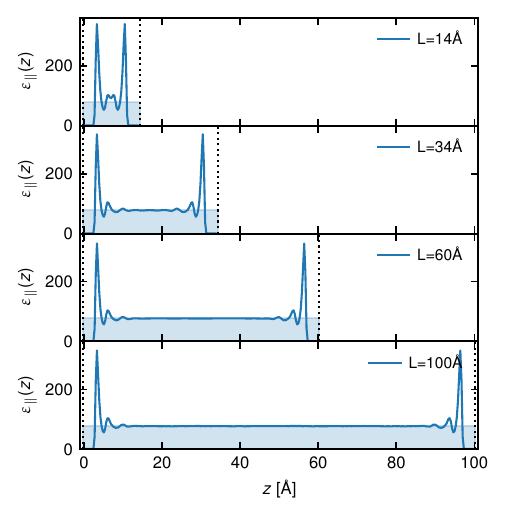}
    \includegraphics[width=0.49\textwidth]{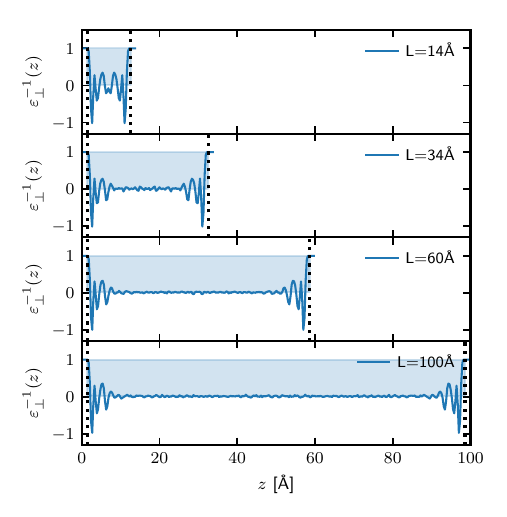}
    \caption{Visualization of the effective dielectric box as determined from effective medium theory for all pore sizes. The dashed lines show the location of the dielectric dividing surface, the shaded, blue region is the equivalent dielectric \enquote{profile} of the box model.}
    \label{fig:effective_lengths}
\end{figure}

\section{Simulation Methods}
\label{sec:methods}

\begin{figure}[ht]
    \centering
    \def\svgwidth{\columnwidth}
    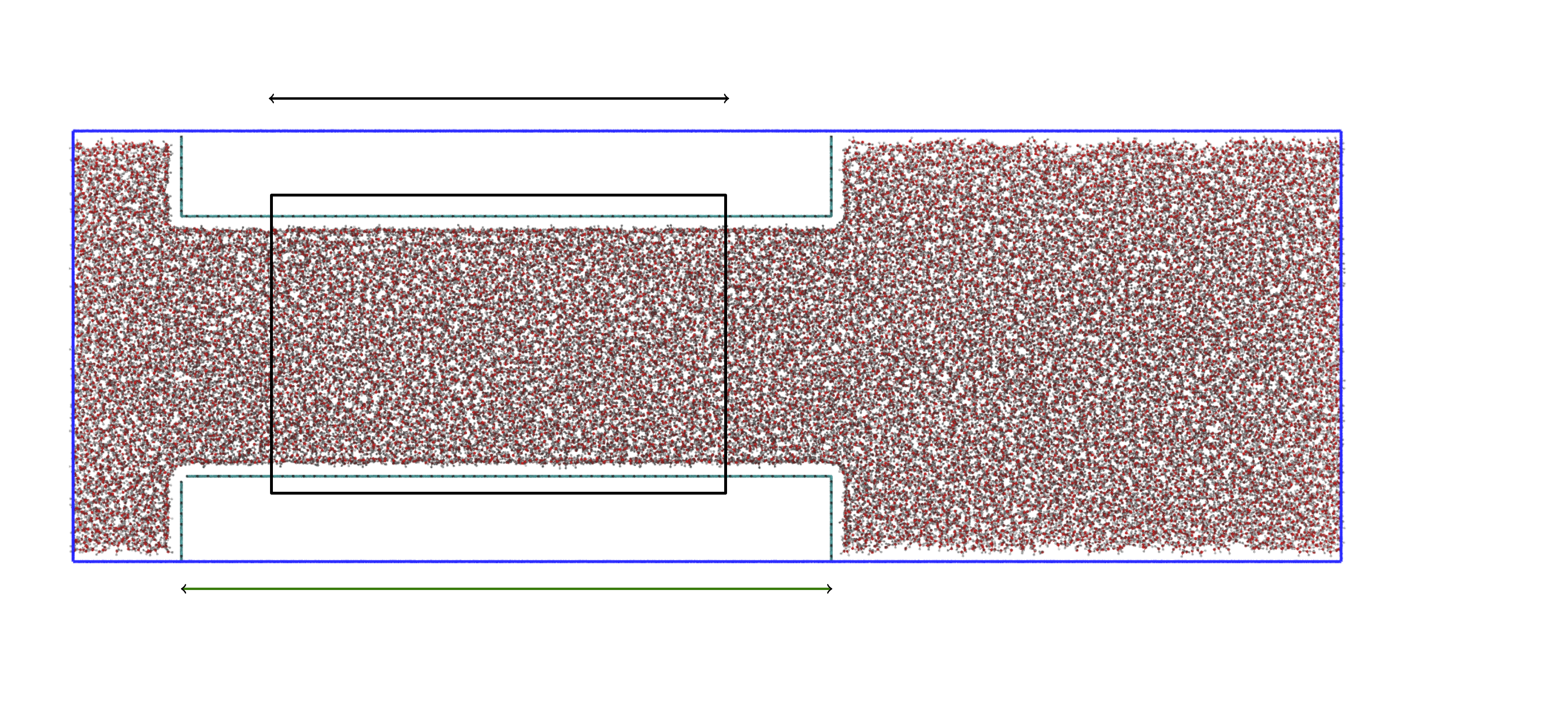

    \caption{Visualization of the piston setup used to determine different water numbers for the main text, which are labeled as SPC/E\(^*\). This snapshot shows the exemplary system of $L = \SI{60}{\angstrom}$. Simulations were carried out with full 3d periodicity.}
    \label{fig:piston_setup}
\end{figure}

Simulations described in this work were performed using both of the simulation frameworks LAMMPS~\cite{thompson22a} (stable release 23 June 2022) and GROMACS~\cite{abraham15a} (release 2021.5) to check for consistency in all calculated physical observables.
Graphene atoms were modeled using the force-field GROMOS 54A7~\cite{schmid11a} and water molecules were simulated using the TIP4P/$\epsilon$~\cite{fuentes-azcatl14a} and SPC/E~\cite{berendsen87a} force-fields.
Following our previous work \cite{loche20a}, a lattice constant of $\SI{1.4}{\angstrom}$ was used for the graphene.
Simulation trajectories were collected for at least $\SI{1}{\micro\second}$, but trajectories were extended until the noise level in the dielectric profiles was acceptable, which for systems with $L \ge \SI{34}{\angstrom}$ was at least $\SI{7}{\micro\second}$ of total simulation time.
Graphene atoms were kept frozen during the trajectory and the water molecules were kept rigid using the SHAKE (LAMMPS) and LINCS (GROMACS) algorithms \cite{hess97a, ryckaertNumericalIntegrationCartesian1977}.
The lateral dimensions of the graphene sheets was fixed to \( \SI{44.28}{\angstrom} \times \SI{42.6084}{\angstrom} \).
The number of waters for a given slab width for the SPC/E model was determined by equlibrating the confined water's chemical potential with its reservoir \cite{loche20a}, cf.\ \cref{tab:water_numbers}, and also used for simulations of the TIP4P/$\epsilon$ model.
Trajectories were recorded every $\SI{1}{\pico\second}$.
All results were calculated using the analysis tool MAICoS, version 0.7.2 \cite{maicos}.
All simulation input files and analysis scripts are freely available at \footnote{\url{https://doi.org/10.18419/DARUS-4317}}.

As is described in the main text, we also investigated the difference in water numbers on dielectric behavior.
To this end, we performed additional simulations of the SPC/E model with a different water number, determined using a piston scheme, which is shown in \cref{fig:piston_setup}.
Here, a system periodic in all directions is compromised of slab and a sufficiently large bulk region is used to flood a channel of given size $L$ while applying a constant pressure laterally to one side of the box.
To achieve this setup in GROMACS, we utilize the Parinello-Rahman barostat in an anisotropic configuration.
The pore size in lateral directions was taken to be 16 by 9 graphene unit cells, with the same graphene lattice parameters as detailed above.
To determine the water number in a region sufficiently far from the edges of the pore, we took the average water number per surface area in a region of length $ H _{\text{slab}}=\SI{50}{\angstrom}$ in the middle of the pore.
The results of this procedure are shown in \cref{tab:water_numbers} in the row labeled SPC/E$^*$.
It was our goal with this procedure to investigate if a different approach that one could have taken will affect the confined water's dielectric behavior.
The main difference between SPC/E and SPC/E$^*$ are the fact that a lateral pressure is used to control densities and that there might be finite size effects due to the periodicity of the system.

\begin{table}
\caption{Number of water molecules used in the simulations shown in this work. The column heading give the separation $L$ between the graphene sheets.
The SPC/E model uses the same number of water molecules as \citeauthor{loche20a} \cite{loche20a}.
For comparability, these numbers were also used for the TIP4P/$\epsilon$ model.
The simulations labelled as SPC/E (*) contain slightly more water molecules obtained from the piston setup, used to investigate how this difference manifests in the dielectric behavior.}
\label{tab:water_numbers}
\begin{tabular}{@{}lllll@{}}
\toprule
Water Model & 1.4 nm     & 3.4 nm     & 6 nm     & 10nm \\ \midrule
SPC/E  &   626      &  1871      &  3479    & 5979      \\
TIP4P/$\epsilon$ &  626      &  1871      &  3479    & 5979     \\
SPC/E$^*$    & 646        &   1910     &  3547    & 6071     \\ \bottomrule
\end{tabular}
\end{table}

\section{Derivation of 3d Formula}
\label{sec:3d_si}
The general derivation for reaction-field boundary conditions can be found in~\cite{stern03a}, we here summarize the derivation of the perpendicular dielectric permittivity profiles for 3d-periodic slab systems with tin-foil boundary conditions.
In this case, the field due to periodic images alters the effective displacement field (also see refs.\ \cite{gekle12a,loche19a,stern03a,neumann83a}),
\begin{equation}
    \label{eq:D_eff}
D _{\perp}^\text{eff} = D_\perp^\text{ext} - \Delta M_\perp/V,
\end{equation}
where $V$ is the entire system's volume and $\Delta M_\perp$ the change in the system's dipole upon application of the external displacement field $D_\perp^\text{ext}$.
If the volume is sufficiently large, as discussed in the main text, the last term becomes negligible and the 2d case is recovered.
In order to derive the Fluctuation-Dissipation relation, eq.\ (33) in the main text, we start with the electrostatic Hamiltonian for a polarization density $m_\perp(z)$ in an external field $E_\perp(z)$, which is given by
\begin{equation}
    \label{eq:hamil_e_field}
H _{\text{el}} = - A \int \dd z\, E_\perp(z) m_\perp (z),
\end{equation}
where we have integrated out the lateral dimensions $x, y$.
Inserting the definition of the effective displacement field, \cref{eq:D_eff}, yields
\begin{equation}
    \label{eq:hamil_d_field}
H _{\text{el}} = - A \int \dd z\, \frac{D_{\perp}^\text{eff} m_\perp (z)}{\epsilon_0} + A \int \dd z\, \frac{m^2_\perp(z)}{\epsilon_0}.
\end{equation}
Within linear response we can linearize eq.\ (30) in the main text, yielding
\begin{equation}
    \label{eq:m_lin}
\Delta m_\perp(z) = D_\perp (1-\epsilon^{-1}_\perp(z)),
\end{equation}
and using $M_\perp = A \int \dd z\, m_\perp(z)$, we can write \cref{eq:D_eff} as
\begin{equation}
    \label{eq:d_eff_reformulated}
D_{\perp}^{\text{eff}} = D_\perp^\text{ext} - \frac{\Delta M_\perp}{V} = D_\perp^\text{eff} - \frac{A}{V} \int \dd z \Delta m_\perp (z) = D_\perp^\text{ext} - \frac{D_\perp^\text{ext}}{L} \int \dd z\, (1-\epsilon^{-1}_\perp (z)) 
= D_\perp^\text{ext} B,
\end{equation}
where we have introduced the average of the inverse dielectric profile,
\begin{equation}
    \label{eq:eps_av}
B :=  \frac{1}{L} \int \dd z \epsilon_\perp^{-1}(z).
\end{equation}
Inserting \cref{eq:d_eff_reformulated} into \cref{eq:hamil_d_field} thus yields
\begin{equation}
    \label{eq:hamil_ref_formulated}
H _{\text{el}} = - A \frac{D_\perp^\text{ext} B}{\epsilon_0} \int \dd z\, m_\perp(z) 
    + A \int \dd z\, \frac{m^2_\perp(z)}{\epsilon_0}.
 = - \frac{D_\perp B}{\epsilon_0} M_\perp + A \int \dd z\, \frac{m^2_\perp(z)}{\epsilon_0}.
\end{equation}

Hence, the derivation of the Fluctuation-Dissipation relation is modified comparing the 2d to the 3d case,
\begin{align}
    \label{eq:def_fluct}
\eval{\pdv{\expval{\Delta m_\perp (z)}}{D_\perp^\text{ext}}}_{D_\perp^\text{ext} = 0} 
    &= \eval{\partial_{D_\perp^\text{ext}} \frac{\sum_i \Delta m_{\perp,i}(z) \exp(-\beta \left[H_{0,i} - \frac{B}{\epsilon_0 } D_\perp^\text{ext} M_{\perp,i} + \dots\right])}{Z}}_{D_\perp^\text{ext} = 0} \nonumber \\
    &= \frac{\beta B}{\epsilon_0 }\left( \expval{m_\perp(z)M_\perp} - \expval{m_\perp(z)}\expval{M_\perp}\right).
\end{align}
with $Z$ being the partition sum, $H_0$ the Hamiltonian of the unperturbed system and the index $i$ referring to the $i$-th configuration.

The latter equation can be related to \cref{eq:eps_av} by making use of eq.\ (30) in the main text, $\eval{\pdv{\expval{\Delta m_\perp(z)}}{D_\perp^\text{ext}}}_{D_\perp^\text{ext} = 0} = (1 - \epsilon^{-1} (z) )$.
Averaging \cref{eq:def_fluct} over $z$ thus gives
\begin{equation}
    \label{eq:integrated}
1 - B = \frac{\beta B}{\epsilon_0 V }\left( \expval{M_\perp^2} - \expval{M_\perp}^2\right) =: \frac{\beta B}{\epsilon_0 V}C_\perp,
\end{equation}
which allows us to solve for $B$, finally yielding
\begin{equation}
    \label{eq:solve_eps_over}
B = \frac{1}{C_\perp \beta/(\epsilon_0 V) + 1 }.
\end{equation}
Solving \cref{eq:def_fluct} for $\epsilon_\perp^{-1}(z)$ by employing eq.\ (30) in the main text thus yields the corresponding expression given in eq.\ (32) of the main text,
\begin{equation}
    \label{eq:3d_eps_perp}
\epsilon_\perp ^{-1}(z) = 1 - \frac{\expval{m_\perp(z)M_\perp} - \expval{m_\perp(z)}\expval{M_\perp}}{\epsilon_0 k _{\text{B}}T + C_\perp/V}.
\end{equation}

\section{Error Estimation for Effective Dielectric Constants}
In order to provide error estimates for the effective dielectric constants and dielectric shifts $\delta^w_\alpha$ we performed a scheme described in the following.
The software package MAICoS provides estimates for the standard error $\Delta \epsilon_\alpha(z)$ of dielectric profiles $\epsilon_\alpha(z)$ per bin.
In the following we will use $\epsilon_\alpha$ as a notational symbol for $\epsilon_\parallel$ and $\epsilon_\perp^{-1}$, respectively, as it drastically simplifies the notation.
In the end, one only has to correct the error estimation of $\epsilon_\perp^{\text{eff}}$ in order to account for the reciprocal nature of $\epsilon_\perp^{-1}(z)$.
In order to perform the numerical integral of this discrete function we use the trapezoidal rule
\begin{equation}
    \label{eq:trapz}
I_\alpha = \sum_{i=1} \frac{\epsilon_\alpha(z_{i+1}) + \epsilon_\alpha(z_i)}{2} \Delta z,
\end{equation}
where $\Delta x$ is the bin width of the dielectric profile $\epsilon_\alpha (z)$.
Following standard error propagation, we estimate the standard error of the numerical integral $I$, $\Delta I$ via
\begin{equation}
    \label{eq:trapz}
\Delta I_\alpha = \sum_{i=1} \frac{\Delta \epsilon_\alpha(z_{i+1}) + \Delta \epsilon_\alpha(z_i)}{2} \Delta z,
\end{equation}
This then allows us to estimate the error on the effective lengths, $\Delta L _\alpha^{\text{eff}}$ via further error propagation, giving us
\begin{equation}
    \label{eq:err_leff}
    L^{\text{eff}}_\alpha = \frac{I_\alpha - L}{\varepsilon^{\text{eff}}_\alpha - 1}\\
\Delta L^{\text{eff}}_\alpha = \left| \frac{1}{\varepsilon^{\text{eff}}_\alpha - 1} \right| \Delta I_\alpha,
\end{equation}
setting $\epsilon_\alpha^{\text{eff}}$ to the average of values measured in bulk (which we define as $\SI{15}{\angstrom}$ distance from pore walls).

We use these estimates to calculate a preliminary estimate of the dielectric interfacial shift via $2\delta^{\text{w}}_\alpha := L_\alpha^{\text{eff}} - L _{\text{w}}$ as described in the main text.
From this we get a improved estimate by averaging all results for a given water model in pores larger than $\SI{1}{\nano\meter}$, giving $\overline{\delta^w_\alpha}$, for which the dielectric shift is expected to be constant. 
In the following, we thus use $L^{\text{eff}} = 2\overline{\delta^{\text{w}}_\alpha} + L _{\text{w}}$ as an improved estimate for the effective dielectric length of all pore sizes.

This then finally lets us calculate the effective dielectric constant and an associated error as
\begin{equation}
    \label{eq:eff_dielectric}
\varepsilon^{\text{eff}}_\alpha = 1 + \frac{I _\alpha- L}{L^{\text{eff}}_\alpha}\\ \Delta\varepsilon^{\text{eff}}_\alpha \approx \left| \frac{1}{L^{\text{eff}}_\alpha} \right| \Delta I_\alpha.
\end{equation}
As stated above, we have to account for the reciprocal nature of $\epsilon_\perp^{-1}(z)$, which we do in the last step, giving us
\begin{equation}
    \label{eq:eff_perp}
\epsilon_\perp^{\text{eff}} = (\epsilon_{\alpha=\perp}^{\text{eff}})^{-1}, \\ \Delta \epsilon_\perp^{\text{eff}} = (\epsilon_{\alpha=\perp}^{\text{eff}})^{-2} \Delta \epsilon_{\alpha=\perp} ^{\text{eff}},
\end{equation}
where $\epsilon_\perp^{\text{eff}}$ is meant to symbolize the value of the effective perpendicular dielectric constant that is reported in the main text and where $\epsilon_{\alpha = \perp}^{\text{eff}}$ refers to the value of the equations above, calculated from $\epsilon_\perp^{-1}(z)$.

\end{document}

%% file: document.tex
\section{Motivation}
\label{sec:motivation}
Nano-porous materials have successfully been used to provide high-capacity energy storage~\cite{oakes13a, mohammadi19a, melnik22a,sun16a},
in catalysis~\cite{kadja22a,wordsworth22a,munoz-santiburcio17a} and for filtration~\cite{simoncelli18a,gravelle16a,priya22a}.
However, the large surface area is not the only factor empowering nano-porous materials for such applications, since the properties of nano-confined fluids such as diffusion/transport or phase behavior can also significantly differ from bulk.
Of particular relevance is the dielectric behavior when fluids are confined to length-scales that approach the size of the constituting molecules.
Simulations have predicted a rather constant, bulk-like dielectric response of water down to confinement lengths of $\sim \SI{1}{nm}$~\cite{bonthuis12a,schlaich16a,loche19a}.
Later experiments, however, have been interpreted in terms of a strong reduction of the dielectric response perpendicular to the surface~\cite{fumagalli18a}.
These experimental observations then spurred further interest and multiple simulation-based studies have since tried to explain in part the anomalous behavior of nano-confined water and other liquids~\cite{stern03a,faraudo04a,ballenegger05a,froltsov07a,nymeyer08a,bonthuis12a,gekle12a,ghoufi12a,zhang13a,parez14a,renou15a,itoh15a,deluca16a,schlaich16a,meneses-juarez18a,loche19a,mondal19a,zhu20a,loche20a,motevaselian20b,motevaselian20a,jalali20a,monet21a,ahmadabadi21a,majumdar21a,hu21a,jalali21a,guo21a,zhu22a,mulpuri23a,borgisDielectricResponseConfined2023,dinpajooh23a,becker24a,tang_scale-dependent_2025}.
Similar effects have been observed in simulations for cylindrical and spherical confinement~\cite{loche19b,schaaf15a,calegariandrade24a}.
However, there are some challenges in the calculation and interpretation of dielectric properties from simulations, stemming mostly from the treatment of the electrostatic boundary conditions~\cite{ballenegger05a,stern03a,bonthuis12a,schlaich16a,gekle13a,loche19a}.
We thus provide here a didactic re-derivation of the concept of dielectric profiles, with particular emphasis on these effects.
This is complemented by a detailed review of the recent literature on dielectric profiles and their interpretation, highlighting possible pitfalls and misunderstandings.

We focus here on the dielectric behavior of water confined between graphene sheets, which has been studied extensively both experimentally~\cite{fumagalli18a,wang24b} and computationally~\cite{loche20a,loche19a,loche19b,loche18a,ruiz-barragan22a,ruiz-barragan20a,dufils24a,monet21a,majumdar21a}.
Yet, there still exists confusion on the correct treatment of the electrostatic boundary conditions in simulations of this system.
Furthermore, a number of different models have been put forward in order to connect the atomistic results to macroscopic observables that can be probed in experiments, which sometimes seem to be at odds with one another.
We extend previous investigations by employing the TIP4P/$\epsilon$ water model, which was optimized to exhibit an accurate dielectric response~\cite{fuentes-azcatl14a} and compare these results to the previously studied SPC/E water model.
Additionally, we investigate the influence of the number of water molecules confined between the graphene sheets on the resulting dielectric profiles, which is usually neglected.
Finally, we discuss approaches to coarse-grain the dielectric profiles using effective medium theory---an approach typically employed to connect to macroscopic measurements~\cite{schlaich16a,loche20a,jalali20a}, but which requires careful interpretation.

\section{Theory of Dielectric Behavior in (Confined) Fluids}
\label{sec:determination_profiles}
To highlight the challenges in dealing with boundary conditions for systems with planar symmetry, we first discuss the dielectric response of bulk fluids and then extend this to anisotropic systems.

\subsection{Theory of Dielectric Response in Bulk Liquids}
\label{sec:bulk_section}
Let us consider the linear polarization response of a liquid subject to an external, homogeneous electrical field of strength $E$.
The resulting polarization is then characterized in terms of the dielectric constant $\epsilon = 1 + \chi$, where $\chi$ is the susceptibility~\cite{feynman11a}.
The relation between the dielectric constant and the liquid's microscopic details has been studied extensively~\cite{lorentz16a,debye35a,fowler35a,onsager36a,vanvleck37a,kirkwood39a,neumann80a,neumann83a,bopp96a,zhang16a}. 
We now sketch the fundamentals of this connection in order to highlight challenges in the determination of dielectric behavior from simulations and theoretical considerations.

To derive an expression for the susceptibility, one considers
the linear response upon application of an external field $\vec{E}$ of the
system's dipole, defined in the absence of a net charge by $\vec{M}=\int
\varrho(\vec{r}) \vec{r} \,\mathrm{d}\vec{r}$.
Here, $\varrho(\vec{r})$ is the instantaneous charge density at position $\vec{r}$ and the integral is over the system's volume.
In the case of $N$ discrete partial charges $q_j$ at positions $\vec{r}_j$ this simplifies to
\begin{equation}
    \label{eq:polarization_def}
\vec{M}\left(\vec{r}_1, \dots, \vec{r}_N\right) = \sum_j \vec{r}_j q_j =: \vec{M}(\Gamma),
\end{equation}
where $\Gamma := (\vec{r}_1, \dots \vec{r}_N)$ is a point in the configurational phase space of the liquid.
In the following, we will label different microstates as $\Gamma_i$.

The linear, dielectric response of the system is then given by
\begin{equation}
    \label{eq:lin_res_def}
    \epsilon = 1 + \frac{1}{\epsilon_0 V} \pdv{\expval{M}}{E}\eval_{E = 0},
\end{equation}
where $E$ is the electric field strength, $\expval{M}$ is the ensemble average of the system's dipole moment parallel to the applied field and $V$ the volume of the fluid under consideration.
The calculation of this expectation value can be performed within the framework of statistical mechanics:
We consider the case of a Hamiltonian perturbed by the external field, $H(\Gamma_i) = H_0(\Gamma_i) - \vec{E}\vec{M}(\Gamma_i)$, where $H_0$ is the system's Hamiltonian in absence of an external field and $-\vec{E}\vec{M}$ is the additional energy  due to the interaction of the system with the external field.
Employing the probability to find the system in microstate $\Gamma_i$ the derivative is expressed as
\begin{equation}
    \label{eq:fluct_diss_bulk}
\eval{\pdv{\expval{M}}{E}}_{E = 0} = \eval{\pdv{}{E} \frac{\sum_i M_i \exp(-\beta H_{0,i} - \vec{E} \vec{M}_i)}{\sum_i \exp(-\beta H_{0,i} - \vec{E} \vec{M}_i)}}_{E = 0},
\end{equation}
where the sums are over all microstates $i$ and we used $M (\Gamma_i) =: M_i$, $H_0(\Gamma_i) =: H_{0,i}$ and where $\beta :=(k _{\text{B}}T)^{-1}$ is the inverse thermal energy. 
The fundamental ingredient of linear response theory is now to expand the exponentials in \cref{eq:fluct_diss_bulk} to first order, yielding
\begin{equation}
    \label{eq:fluct_diss_bulk2}
\pdv{\expval{M}}{E} = \beta \left( \expval{M^2}-\expval{M}^2 \right) =: \beta C,
\end{equation}
where we introduced $C$ for the variance of the system's dipole.

\begin{figure}
    {\Large
    \centering
    \def\svgwidth{\columnwidth}
    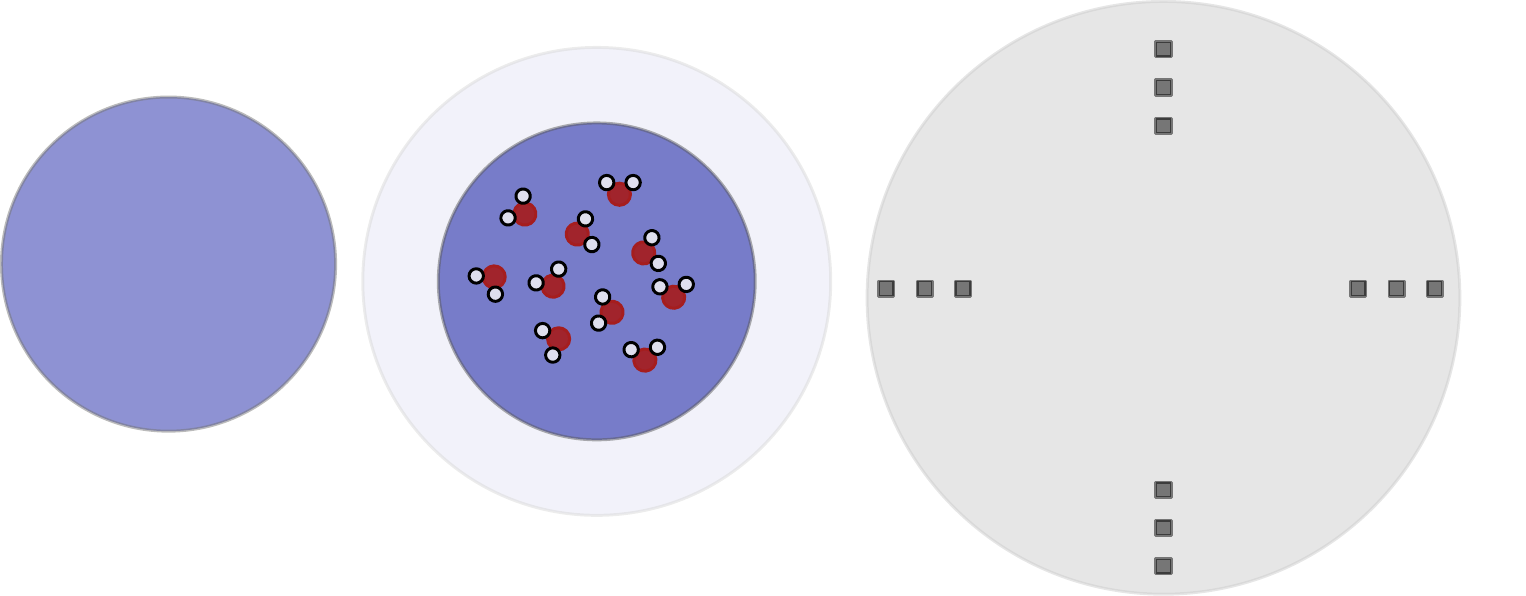

    \caption{Schematic illustration of typically employed boundary conditions for dielectric properties of a bulk fluid in simulations and theory. (a) Liquid droplet immersed in vacuum and (b) in a dielectric continuum with the same properties as the fluid. (c) Periodic boundary conditions. For methods such as the Ewald approach to solve the Coulomb sum, a dielectric boundary condition at infinity, $\epsilon_\infty$, needs to be specified.}
    \label{fig:kirk_f_claus}
}
\end{figure}

Crucially, this derivation does not take into account any boundary condition, it rather assumes an infinite homogeneous sample of the fluid.
In contrast, the exact response function in \cref{eq:lin_res_def} depends on the boundary condition encountered in simulations or theoretical models (but also in experiments), as we discuss now in detail.
They are typically accounted for by either considering an isolated sample of molecules (Clausius--Mossotti \cite{neumann83a} and Kirkwood--Fröhlich approach~\cite{kirkwood39a}) or by employing periodic boundary conditions (Neumann approach~\cite{neumann83a}), see \cref{fig:kirk_f_claus}.

The simplest assumption considers an isolated, spherical sample in vacuum, leading to an equation which relates the dipole moment of a small spherical sample of volume $V$ to the dielectric constant of the bulk fluid~\cite{neumann83a}, see \cref{fig:kirk_f_claus}(a):
\begin{equation}
    \label{eq:clausius_mossotti}
    \frac{\epsilon - 1}{\epsilon + 2} = \frac{1}{9 \epsilon_0} \frac{C _{\text{isolated}}}{ V k _{\text{B}}T},
\end{equation}
where $C _{\text{isolated}}$ refers to the fact that the dipole variance is calculated from an isolated sample of the fluid.
Due to the similarity of the l.h.s.\ of \cref{eq:clausius_mossotti} with the original Clausius--Mosotti relation~\cite{rysselberghe32b,mossotti50a,clausius79a}, this is typically referred to as the Clausius--Mosotti approach.

Another approach is to calculate the dielectric constant from a sub-sample of a liquid surrounded by a homogeneous medium of the same dielectric constant as illustrated in \cref{fig:kirk_f_claus}(b).
Since the sample induces a polarization in the surrounding medium of dielectric constant $\epsilon_\mathrm{RF}$, this in turn creates a reaction field in the sample, leading to the Kirkwood--Fröhlich equation~\cite{kirkwood39a}
\begin{equation}
    \label{eq:kirkwood_frohlich}
    \frac{\epsilon -1}{\epsilon + 2}\left[ 1+ \frac{\epsilon -1}{\epsilon + 2} \frac{2(\epsilon_{\text{RF}} -1) }{2\epsilon_{\text{RF}} + 1} \right]^{-1} = \frac{1}{9\epsilon_0} \frac{\expval{C_{\text{RF}} ^2}}{V k_{\text{B}} T},
\end{equation}
with the subscript highlighting that the dipole variance $C _{\text{RF}}$ is calculated for a sample under reaction-field boundary conditions.

A comparison of \cref{eq:clausius_mossotti,eq:kirkwood_frohlich} shows that the boundary conditions (isolated sample vs.\ reaction-field) matter greatly for the determination of the dielectric response.
Still, Clausius-Mossotti and Kirkwood-Fröhlich-like approaches have been employed outside their validity for fluids in planar confinement~\cite{ryazanov06a,sato18a}, which will be discussed below.

Given the long-ranged nature of the Coulomb potential and to improve numerical efficiency, most modern simulations mimic an infinite system by considering periodic boundary conditions as indicated in \cref{fig:kirk_f_claus}(c), reducing artifacts from cutoff electrostatics~\cite{sagui99a,borwein85a,schreiber92a,york95a}.
Electrostatic forces are usually calculated via some variant of an Ewald-sum method in order to efficiently calculate the conditionally convergent sum over the periodic images~\cite{ewald21a,hockney88a,deserno98a,deserno98b}.
This approach is obviously distinct from both Kirkwood-Fröhlich and Clausius-Mossotti-type approaches leading to \cref{eq:kirkwood_frohlich,eq:clausius_mossotti} and requires yet other fluctuation relations~\cite{neumann80a,neumann83a}.
Here again, careful considerations of the electrostatic boundary conditions are required to correctly capture the significance of the system's dipole term~\cite{deleeuw97a}.
Most common simulations employ the so-called \emph{tin-foil} boundary conditions, corresponding to an external medium with $\epsilon_{\text{RF}} = \infty$~\cite{smith81b,roberts95a,deleeuw97a,boresch97a}.
Any polarization of the system does not induce a reaction field in this case and one obtains~\cite{neumann80a}
\begin{equation}
    \label{eq:tin_foil_homogeneous}
    \epsilon = 1 + \frac{1}{3\epsilon_0} \frac{C _{\infty,\text{TF}}}{ V k _{\text{B}}T}.
\end{equation}
In contrast, if a reaction-field-like boundary condition is used, i.e.\ $\epsilon_\infty = \epsilon$, the reaction-field is non-zero and the dielectric constant is given by
\begin{equation}
    \label{eq:reaction_field_ewald_homogeneous}
    \frac{(2\epsilon + 1)(\epsilon - 1)}{9\epsilon} = \frac{1}{9\epsilon_0} \frac{C _{\infty, \text{RF}}}{V k _{\text{B}} T},
\end{equation}
where we again used the notation $C _{\infty,\text{TF}}$ and $C _{\infty, \text{RF}}$ to highlight that the equations are only correct provided the dipole variance is calculated from simulations with periodic images and either tin-foil or reaction-field dielectric boundary conditions.

While the outlined relations are well-established, it is useful repeating them in the present context, as there seems to be considerable confusion on the applied boundary conditions in literature and corresponding simulations of fluids at interfaces---as discussed below---clearly have to be corrected similarly and with great care.
For example, bulk reaction-field methods presented in this section have been employed to investigate the effective dielectric constant in spherical cavities as a function of their radius~\cite{senapati01a}, yet the observed reduction with respect to bulk sensitively depends on the definition of the probe volume $V$.

What we have discussed above is the local, static dielectric response.
However, the microscopic details of the dielectric response in liquids have been elucidated in more detail.
Starting with works by Lorentz and Onsager among others that aimed to understand the dielectric response as a function of the polarizability of individual molecules~\cite{lorentz16a,debye35a,fowler35a,onsager36a,vanvleck37a}, it was realized by Kirkwood that correlations between molecules are crucial to be included~\cite{kirkwood39a}.
This extension of Onsager's theory, accounting for the enhancement due to molecular correlations, is commonly quantified in terms of Kirkwood's $g$-factor, which can be calculated systematically from simulations as a function of the distance from a molecule~\cite{zhang16a}.
Similarly, the non-local dielectric response of water has been investigated in simulations~\cite{bopp96a}.
As we discuss next, the non-locality of the dielectric response in anisotropic systems can be incorporated into an effective local response.

\subsection{Dielectric Profiles in Planar Geometry}
\begin{figure}[h]
    \centering
    \includegraphics[width=\linewidth]{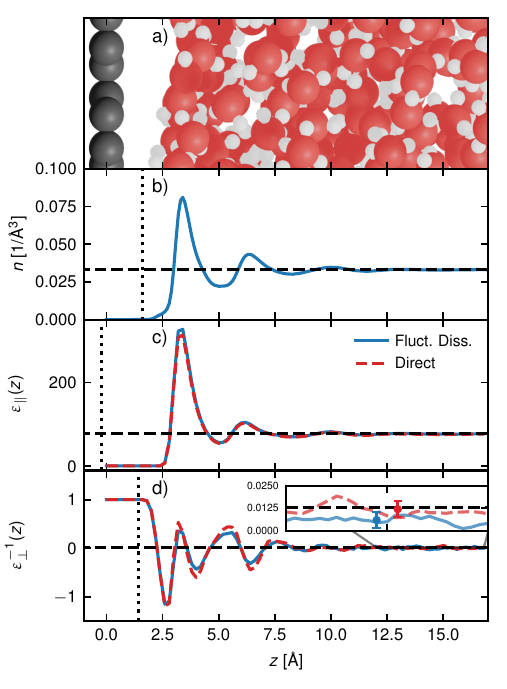}
    \caption{Snapshot of the simulation system and representative observables discussed in this work. Data shown are obtained using the TIP4P/$\epsilon$ model for water confined between insulating graphene sheets separated by $\SI{34}{\angstrom}$. The position $z = \SI{0}{\angstrom}$ denotes the location of the carbon atoms.
    (a) Simulation snapshot showing carbon as gray, oxygen as red and hydrogen as white spheres.
    (b) Number density profile $n(z)$ of water molecules. The dashed line shows the bulk density $n_\mathrm{bulk}=\SI{0.033}{\angstrom^{-3}}$ and the dotted line the location of the Gibbs dividing surface. (c) Parallel dielectric profile. The dashed line shows the dielectric constant $\epsilon = 79$ of the water model at $\SI{300}{\kelvin}$.
    (d) Inverse perpendicular dielectric profile. The dashed line shows the corresponding bulk value $\epsilon^{-1} = 1/79$.
    In (c) and (d), the blue lines show results obtained from equilibrium fluctuations of the dipole density [\cref{eq:par_fluct_diss,eq:perp_fluct_diss}] and the red via the \enquote{direct} route [\cref{eq:lin_res_par,eq:lin_res_perp}] as explained in the text, using $E_\parallel = \SI{0.05}{\volt\per\nano\meter}$ and $D_\perp/\epsilon_0 = \SI{0.2}{\volt\per\nano\meter}$, respectively.
    The dotted lines in (c) and (d) denote the corresponding locations of the dielectric dividing surfaces. 
    The inset in (d) also shows the mean over the profiles for $z > \SI{12}{\angstrom}$ as round markers with corresponding error bars.
    }
    \label{fig:example_epsilon_profs}
\end{figure}

The general linear response relation in a heterogeneous system is expected to be anisotropic and non-local.
If we are in a linear response regime, the dielectric response of the displacement field $\Delta \vec{D} (\vec{r})$ due to an external field $\Delta \vec{E}(\vec{r})$ is thus given by \cite{kornyshev86a}
\begin{equation}
    \label{eq:nlin_aniso_dielectric}
\Delta \vec{D}(\vec{r}) = \epsilon_0 \int \epsilon(\vec{r}, \vec{r}') \Delta \vec{E} (\vec{r}') \mathrm{d}\vec{r}',
\end{equation}
where $\epsilon(\vec{r}, \vec{r}')$ is the dielectric tensor giving the non-local response of $\Delta \vec{D}(\vec{r})$ at position $\vec{r}$ due to a field at $\vec{r}'$.

Without loss of generality, we now assume that the planar system is confined in the $z$-direction.
The Maxwell-equations must hold, hence in the absence of a magnetic field
\begin{equation}
    \label{eq:rot_e}
\nabla \times \vec{E} = 0.
\end{equation}
Due to translational invariance parallel to the surface, the field only varies in $z$-direction, i.e.\
the parallel field $E_\parallel$ is constant throughout the box.
We can thus re-write \cref{eq:nlin_aniso_dielectric} for the parallel components $\epsilon_\parallel$ of the dielectric tensor~\cite{stern03a,bonthuis12b}:
\begin{equation}
    \label{eq:par_eps_start}
\Delta D _\parallel (\vec{r}) = \epsilon_0 \Delta E_\parallel \int \mathrm{d}\vec{r}' \epsilon_\parallel(\vec{r}, \vec{r}'). 
\end{equation}
Next, the translational invariance allows us to integrate \cref{eq:par_eps_start} over the lateral directions and we thus arrive at
\begin{equation}
    \label{eq:par_eps_def}
\Delta D_\parallel (z) = \epsilon_0 \Delta E_\parallel \int \mathrm{d}z' \epsilon_\parallel(z, z') =: \epsilon_0 \epsilon_\parallel(z) \Delta E_\parallel,
\end{equation}
where the last step defines the local dielectric profile $\epsilon_\parallel(z)$.
Importantly, this definition follows directly from the non-local response and is exact, contrary to a common misconception in the literature~\cite{cox22a}.

Determining the second independent component follows similar lines:
We start with the inverse dielectric response
\begin{equation}
\Delta \vec{E} (\vec{r}) = \epsilon_0^{-1} \int \mathrm{d}z' \epsilon^{-1} (\vec{r}, \vec{r}') \Delta \vec{D}(\vec{r}'),
\label{eq:perp_eps_start}
\end{equation}
where $\epsilon^{-1}(\vec{r}, \vec{r}')$ is the functional inverse of the non-local dielectric tensor.
If there are no free charges in the system, according to Maxwell's relations
\begin{equation}
    \label{eq:free_charge_macroscopic_maxwell}
\nabla \vec{D}(\vec{r}) = 0,
\end{equation}
and thus the perpendicular component of the displacement field $D_\bot$ is constant.
This is an important point, which often is either simply ignored~\cite{hu21a} or not discussed~\cite{jalali21a, coelho_interplay_2024, choganiEffectSalinityDielectric2024, fuentes-azcatl_dielectric_2025}.
Strictly, \cref{eq:free_charge_macroscopic_maxwell} only allows for the calculation of
a local static perpendicular dielectric profile if there
are no ions present in the fluid.
Following the argumentation in the derivation of $\epsilon_\parallel(z)$ above to integrate out the lateral degrees of freedom, the local dielectric response profile is obtained,
\begin{equation}
    \label{eq:perp_eps_def}
\Delta E_\bot (z) = \frac{\Delta D_\bot}{\epsilon_0} \int \dd z' \epsilon^{-1}_\perp (z, z') =: \frac{\Delta D_\bot}{\epsilon_0} \epsilon^{-1}_\bot(z),
\end{equation}
where we have introduced the local dielectric profile $\epsilon_\bot^{-1} (z)$,
which is position-dependent but includes non-local polarization effects and follows directly from $\epsilon (\vec{r}, \vec{r}')$.

A route for systems containing free charges follows from considering the generalized, frequency-dependent dielectric response
$\Sigma(\omega)=\varepsilon(\omega) - 1 +  (4\pi i \sigma(\omega)) / \omega$, where $\sigma$ denotes the ionic AC conductivity \cite{bottcher74a}.
As first discussed by Caillol et al.\ \cite{caillol_theoretical_1986}, the susceptibility of an aqueous solution of ions consists of three separate terms corresponding to the solvent, the ion-solvent cross-correlations, and the ionic contribution \cite{rinneDissectingIonspecificDielectric2014}.
The static dielectric constant for the solution can be defined from the low frequency behavior,
$\varepsilon - 1 = \lim_{\omega \to 0} \left[ \Sigma(\omega) - (4\pi i \sigma(0)) / \omega\right]$.
Since in the direction perpendicular to the planar interface the steady state current must vanish, $\sigma(0) = 0$, this allows to extract information on the perpendicular dielectric profile as follows.
If the displacement field $D_\perp (z)$ varies only slowly relative to the dielectric profile, the integral on the right-hand side of \cref{eq:perp_eps_start} directly yields $\varepsilon_{\perp}^{-1} (z)$
\footnote{Partial integration of the laterally averaged version of \cref{eq:perp_eps_start} yields
\begin{align*}
        \Delta E_\perp (z) &= \varepsilon_0^{-1} \int \mathrm{d}^\prime \, \varepsilon_\perp^{-1}(z,z^\prime) \Delta D_\perp(z^\prime) \\
        &= \varepsilon_0^{-1} \varepsilon_\perp^{-1}(z) \Delta D_\perp(z) - \int \mathrm{d}z^\prime \, \varepsilon_\perp^{-1}(z^\prime) \dv{\Delta D_\perp(z^\prime)}{z^\prime}.
    \end{align*}
    If $D_\perp(z)$ varies slowly with $z$, the second term can be neglected and we recover \cref{eq:perp_eps_def}.
    \label{footnote:slow_varying}
}.
This approximation is expected to hold at low salt concentration and low surface charge density.
In that case, \cref{eq:perp_eps_def} is recovered and the ion distribution can be rationalized in term of a modified Poisson--Boltzmann model \cite{bonthuis12a, schlaich19a, gardre_dielectric_2025}.
Further complication arises if ionic liquids are considered, since in that case both the dielectric constant $\varepsilon(\omega)$ and the conductivity $\sigma(\omega)$ are due to the very same molecule, prohibiting the identification of the solvent dielectric response $\varepsilon_\mathrm{sol}$ \cite{schroder_computation_2008}.

Examplary profiles both of $\epsilon_\parallel(z)$ and $\epsilon_\perp^{-1}(z)$ together with a simulation snapshot and the water number density profile are shown in \cref{fig:example_epsilon_profs} for 1871 TIP4P/\(\epsilon\) water molecules confined between uncharged and positionally fixed graphene sheets of dimensions $\SI{44.28}{\angstrom}\times\SI{42.61}{\angstrom}$ separated by $\SI{34}{\angstrom}$. 
All simulation details are summarized in section II of the Supplementary Information~\cite{suppa} and details on the calculation of the profiles will be presented in the next section. 
Strong water depletion from the hydrophobic surface can be observed in \cref{fig:example_epsilon_profs}(a) and (b), followed by marked density oscillations which is characteristic for a stiff surface.
The parallel dielectric profile in \cref{fig:example_epsilon_profs}(c) is roughly proportional to the number density, as would only be expected for non-interacting dilute polar particles---however, the discrepancies in the peak positions and oscillation periods already hint that this simple picture is not sufficient to describe the dielectric profile of liquid water~\cite{bonthuis12a}. 
The inverse perpendicular response shown in \cref{fig:example_epsilon_profs}(d) crosses zero multiple times, indicating singularities in $\varepsilon_\perp(z)$ and spatial regions with negative response.
These zero crossings give rise to local minima in the corresponding electrostatic potential, which in turn underpin rich interfacial phenomena such as overscreening.
A detailed discussion of these effects can be found elsewhere~\cite{gonella21a,becker24a,hedley_what_2025}.

\section{Dielectric Profiles from Molecular Simulations}
\subsection{Calculating Dielectric Profiles the \enquote{direct way}}
The most intuitive and straightforward way to obtain dielectric profiles is measuring the system's response to an external field and from this retrieve the dielectric profiles according to \cref{eq:par_eps_def,eq:perp_eps_def}.
To this end, a simulation at zero external field strength is performed and compared to results for a finite field.
From these data, the linear response in $\Delta D_\parallel (z)$ and $\Delta E_\perp (z)$ is obtained, respectively.
The definition of the displacement field in polarizable media is~\cite{landau75b}
\begin{equation}
    \label{eq:displacement_def}
    \vec{D}(\vec{r}) = \epsilon_0 \vec{E}(\vec{r}) + \vec{P}(\vec{r}),
\end{equation}
where $\vec{P}(\vec{r})$ is the polarization density at location $\vec{r}$.
We can now make use of the fact that $E_\parallel$ and $D_\perp$ are constant and apply a charge-proportional, spatially constant external force in parallel (perpendicular) direction that correspond to a $E_\parallel$ ($D_\perp$) field~\cite{bonthuis12a,loche19a}.
Such simulations with external fields acting on charges or charge densities can readily be performed with modern simulation packages.
To remain within the linear response regime, the external fields have to remain at sufficiently small strengths, which needs to be addressed carefully~\cite{bonthuis12a,wolde-kidan21a}.

The excess polarization field $\Delta \vec{P} (\vec{r})$ due to an applied field is related to the instantaneous excess polarization density $\Delta \vec{m}(\vec{r})$ in a simulation via the ensemble average
\begin{equation}
    \label{eq:polarization_instant}
    \Delta \vec{P}(\vec{r}) = \expval{\Delta \vec{m}(\vec{r})},
\end{equation}
where the excess refers to the response in polarization to due an externally applied field, i.e.,
\begin{equation}
    \label{eq:response_m}
\expval{\Delta \vec{m} (\vec{r})} := \expval{\vec{m}(\vec{r})} - \expval{\vec{m}_0 (\vec{r})}.
\end{equation}
Here, $\vec{m}(\vec{r})$ denotes the polarization density in a simulation with an external field and $\vec{m}_0(\vec{r})$ to that of a simulation without an external field, respectively.
Note that $\vec{m}(\vec{r})$ refers to the complete polarization, including all terms beyond the dipole.
Due to translational invariance in the planar system we average over lateral dimensions and thus obtain $\Delta \vec{m}(z)$.
Note that the system dipole is connected to the polarization density through spatial integration, $\vec{M} = \int \dd \vec{r}\, \vec{m}(\vec{r})$.

Relating \cref{eq:response_m,eq:displacement_def} to the definitions of the perpendicular and parallel dielectric profiles given in \cref{eq:perp_eps_def,eq:par_eps_def} yields
\begin{equation}
    \label{eq:lin_res_par}
    \epsilon_\parallel (z) = \frac{\epsilon_0 E_\parallel + \expval{m_\parallel (z)} - \expval{m_{0,\parallel} (z)}}{\epsilon_0 E_\parallel} 
\end{equation}
and
\begin{equation}
    \label{eq:lin_res_perp}
    \epsilon_\perp^{-1}(z) = \frac{D_\perp - \expval{m_\perp (z)} + \expval{m_{0, \perp} (z)}}{D_\perp}.
\end{equation}
Hence, in practice, the polarization densities $m_\parallel$ and $m_\perp$ have to be calculated from simulation trajectories as follows.

\emph{Parallel polarization density $m_\parallel (z)$:}
In order to derive a general expression for the polarization density, we assume in the following that the molecules under study are comprised of point charges, which allows calculating $m_\parallel(z)$ via the \emph{virtual cutting method}~\cite{bonthuis12a}.
\citeauthor{stern03a} proposed to decompose the molecular polarization into dipole moments along covalent bonds~\cite{stern03a}\hlork{, which also allows to explicitely include electronic polarizability}~\cite{beckerInterfacialVsConfinement2024}.
Another alternative consists in expanding the molecular polarizations in terms of multipoles~\cite{bonthuis12a}, which can also be derived from electronic structure calculations in terms of maximally localized Wannier functions~\cite{marzari12a}.
\Citeauthor{ruiz-barragan20a} have employed this method in ab-initio simulations of liquid water in graphene slits limited to the effective molecular dipole moment~\cite{ruiz-barragan20a}, which yields the leading contribution to the parallel dielectric profile~\cite{bonthuis12a}.

\begin{figure}[ht]
    \centering
    \def\svgwidth{\columnwidth}
    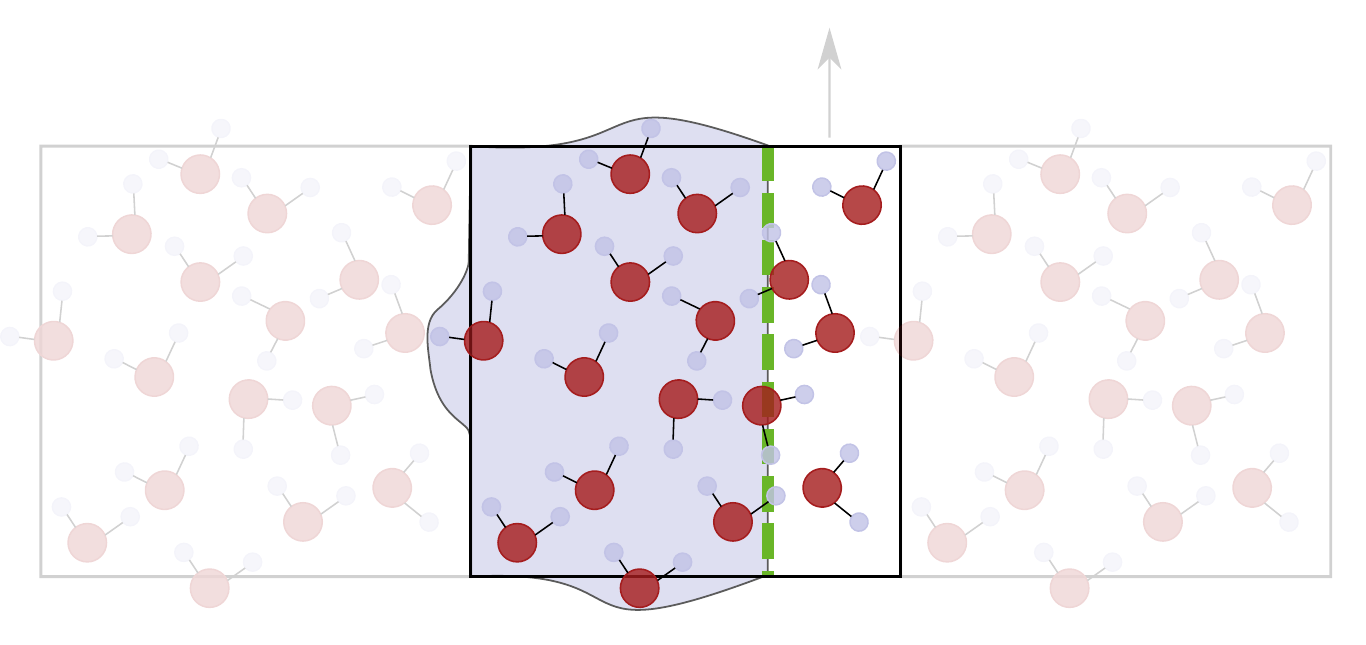

    \caption{Schematic illustration of the virtual cutting method to determine $m_\parallel(z)$. The highlighted box in the middle represents the primary simulation domain. The blue shaded area contains the atom's charges $\rho_{\text{cut}}$, over which the integral is performed in order to arrive at the surface charge density along the green line as a function of the $z$-coordinate. Multiple cuts (different positions of the cutting plane $a_\text{cut}$) are averaged over to obtain an estimate of the parallel polarization density $m_\parallel(z)$.
    }
    \label{fig:virtual_cutting}
\end{figure}

In the virtual cutting approach, planes perpendicular to the applied lateral field are introduced, and the surface charges due to water molecules which are cut across these planes are calculated and averaged over (see \cref{fig:virtual_cutting}).
The $z$-position dependent surface charge $\sigma(z)$ follows from the monopole density $\rho _{\text{cut}}(a, z)$ of the molecules inside the plane (gray region in \cref{fig:virtual_cutting}) and is related to the polarization density as $\sigma (z) = - m_\parallel (z)$~\cite{bonthuis12a,schlaich16a}, i.e.\ it is obtained by calculating
\begin{equation}
    \label{eq:virtual_cut}
    -m_\parallel (z) = \sigma (z) = \int_{-\infty}^{a _{\text{cut}}} \dd a \rho _{\text{cut}}(a, z),
\end{equation}
where $a$ denotes any direction in the plane parallel to the surface (usually the $x$ and $y$ axes are chosen).
Technically, this corresponds to counting the total charge in the cut-volume depending on the $z$ position.
Crucially, one has to take care that in the calculation of $\sigma(z)$, molecules are made \enquote{whole}, i.e.\ not split over periodic boundaries, which requires prudent treatment of MD trajectories as offered by specialized simulation analysis packages~\cite{maicos}.
\hlork{This also reveals that the virtual cutting method is only applicable to neutral molecules, since otherwise no unambiguous definition of the neutral integration region---cf.\ left site of the shaded region in \cref{fig:virtual_cutting}---exists.
Consequently, parallel dielectric profiles cannot be obtained straightforwardly from electronic structure calculations using \cref{eq:virtual_cut}.
Solutions involve using the effective molecular dipole moment based on the maximally localized Wannier functions~\cite{ruiz-barragan20a} or, more generally, to explicitly including the electronic polarizability, which requires prudent treatment of the periodic boundary conditions in terms of the Berry phase position operator~\cite{beckerInterfacialVsConfinement2024}.}

\emph{Perpendicular polarization density $m_\perp (z)$:}
Integration of the total charge density $\rho$ according to the first Maxwell relation directly yields the perpendicular electric field,
\begin{equation}
    \label{eq:electric_field}
    E_\perp (z) = E_\perp (0) + \int_0 ^z \dd z' \frac{\rho(z')}{\epsilon_0} .
\end{equation}
Since the displacement field is constant throughout the simulation box, combining this with \cref{eq:displacement_def} directly yields
\begin{equation}
    \label{eq:location_dep_pol_field}
m_\perp (z) = - \int_0 ^z \dd z' \rho (z')
\end{equation}
for the perpendicular polarization density, which can easily be calculated from simulation trajectories by means of numerical integration.
\hlork{Importantly, this approach also is applicable in electronic structure calculations in two-dimensions \cite{deissenbeck21a} since the total charge density $\rho$ is directly accessible and zero outside of the simulation domain, which allows to unambiguously define the integration limits in \cref{eq:location_dep_pol_field}.}

Dielectric profiles obtained from applied field simulations are shown in \cref{fig:example_epsilon_profs}(c) and (d) as red lines and clearly agree with the results obtained from equilibrium fluctuations derived in the next section [blue lines in \cref{fig:example_epsilon_profs}(c) and (d)].
Importantly, whether the finite field applied to calculate $m_\parallel$ and $m_\perp$ still is within the linear response limit has to be carefully evaluated~\cite{bonthuis12a}.
The corresponding strength is usually less in the parallel than in the perpendicular direction, where we find---for the system studied in \cref{fig:example_epsilon_profs}---$E_\parallel = \SI{0.05}{\volt\per\nano\meter}$ and $D_\perp/\epsilon_0 = \SI{0.2}{\volt\per\nano\meter}$ to be sufficiently small to be within linear response and at the same time strong enough to significantly polarize the water to gain reasonable statistics within the simulation.

Last, we note that---alternatively to applying a field directly as force acting on charges---an external field can also be applied using metallic boundary conditions on the confining planar walls~\cite{deissenbeck23b}.
However, great care has to be taken in the nature of the boundary conditions, as one wishes to simulate at a constant displacement field $D_\perp$, which corresponds to open circuit conditions in the electrode (constant charge per conducting electrode).
Discussing this is beyond the scope of this work and we refer to refs.~\cite{deissenbeck21a,deissenbeck23b,dufils24a}.
We limit the discussion in the remainder of this work to inert walls and the corresponding boundary conditions encountered in simulations and present a detailed analysis of dielectric properties at metallic interfaces in a separate manuscript~\cite{stark24b}.

\subsection{Dielectric Profiles from Equilibrium Fluctuations}
\label{sec:dielectric_profiles}

In some cases, the linear response limit is challenging to reach with externally applied fields, e.g.\ systems with flexible bonds, where linear response is not applicable even for small fields (due to the intramolecular polarizability), or systems with soft interfaces~\cite{schlaich16a,wolde-kidan21a}.
However, dielectric profiles can also be obtained from equilibrium fluctuations analogous to the derivation presented above for homogeneous, isotropic fluids.
This approach is advantageous in that it allows the extraction of dielectric profiles from simulations directly, without the need of performing additional simulations with an explicit, external field.
The fluctuations of the local polarization fully determine the dielectric response of the confined system, provided enough statistics is available to evaluate the variance.

\paragraph{Parallel Dielectric Profile:}
We start with the dielectric response of a nano-confined liquid parallel to the surface and will derive the formalism for the perpendicular response below.
\Cref{eq:par_eps_def} gives us the exact relation in the linear response limit:
\begin{equation}
    \label{eq:parallel_comb_instant_m}
 \Delta D_\parallel (z) = \epsilon_0 \epsilon_\parallel (z) \Delta E_\parallel = \epsilon_0 \Delta E_\parallel + \expval{\Delta m_\parallel(z)},
\end{equation}
which directly yields
\begin{equation}
    \label{eq:eps_fluct_def_par}
\epsilon_\parallel (z) = 1 + \frac{\expval{\Delta m_\parallel (z)}}{\epsilon_0 \Delta E _\parallel}.
\end{equation}
In complete analogy to the derivation in bulk, the linear response of the polarization density is given by
\begin{equation}
    \label{eq:delta_m_lin_res}
\expval{\Delta m_\parallel (z)} = \eval{\pdv{\expval{m_\parallel (z)}}{\Delta E_\parallel}}_{\Delta E_\parallel = 0} \Delta E_\parallel,
\end{equation}
%
%
%
%
For a parallel applied field we can write the perturbed Hamiltonian as $H = H_0 - E_\parallel M_\parallel$ and the expectation value of the derivative can be obtained from the corresponding microstates as in \cref{eq:fluct_diss_bulk}.
Performing the linearization of the exponentials as in \cref{eq:fluct_diss_bulk2} yields
\begin{equation}
    \label{eq:fluct_diss}
    \pdv{\expval{m_\parallel (z)}}{E_\parallel}\eval_{\Delta E_\parallel = 0} = \beta \left(  \expval{m_\parallel  (z) M_\parallel} - \expval{m_\parallel (z)} \expval{M_\parallel}\right).
\end{equation}
Thus, the parallel dielectric profile follows from equilibrium fluctuations of the local parallel polarization density,
\begin{equation}
    \label{eq:par_fluct_diss}
\epsilon_\parallel (z) = 1 + \frac{\expval{m_\parallel (z) M_\parallel} - \expval{m_\parallel (z)} \expval{M_\parallel}}{\epsilon_0 k _{\text{B}}T}.
\end{equation}

\paragraph{Perpendicular Dielectric Profile:}
For the perpendicular response we again make use of the constant displacement field in perpendicular direction and start with \cref{eq:perp_eps_def}.
The linear response relation for the perpendicular field is
\begin{equation}
    \epsilon_0 \Delta E_\perp(z) = \epsilon_\bot^{-1} (z) \Delta D_\perp = \Delta D_\perp - \expval{m_\bot(z)},
\end{equation}
and substituting the analogue relation to \cref{eq:delta_m_lin_res} gives
\begin{equation}
    \label{eq:eps_perp_fluct_i}
\epsilon_\bot^{-1} (z) = 1 - \eval{\pdv{\expval{m_\bot}}{D_\bot}}_{D_\perp = 0}.
\end{equation}

For a strictly two-dimensional periodic system the Hamiltonian reads, $H = H_0 - \frac{D_\perp M_\perp}{\epsilon_0}$\cite{bonthuis12a}. 
Calculating the expectation value to linear order and performing the derivative in \cref{eq:eps_perp_fluct_i} then yields the perpendicular component of the dielectric profile,
\begin{equation}
    \label{eq:perp_fluct_diss}
    \epsilon_\perp ^{-1} (z) = 1 - \frac{\expval{m_\perp(z) M_\perp} - \expval{m_\perp(z)} \expval{M_\perp}}{\epsilon_0 k_{\text{B}}T} .
\end{equation}

\subsection{Boundary Conditions in Molecular Simulations}
As already discussed in detail in \cref{sec:bulk_section}, electrostatic boundary conditions have to be treated carefully in the context of the dielectric response of fluids.
Since nowadays most molecular simulation packages are capable of treating electrostatic interactions via Ewald-sum type approaches, we limit the discussion here to simulations with periodic boundary conditions (pbc).
Solving electrostatics in pbc is ambiguous, as the series of Coulomb energies is conditionally convergent~\cite{ewald21a,hockney88a,deserno98a,deserno98b,smith81b}.
This ambiguity can be reconciled by setting the dielectric boundary conditions at infinity, $\epsilon_\infty$ and by considering a shape dependent term~\cite{smith81b,deleeuw97a}, which fixes the order of summation.
Typically, a spherical summation order is chosen and the dielectric boundary conditions at infinity are either set to $\epsilon_\infty = \infty$, commonly referred to as \emph{tin-foil} boundary conditions, or set to some finite value $\epsilon_\infty$.
The \emph{tin-foil} boundary conditions are much more common, given that other boundary conditions lead to often unwanted behavior, such as forces acting against system dipoles~\cite{deleeuw97a}.


For planar slit pores, one would usually like to employ 2d-pbc simulations.
However, calculating 2d-Ewald sums and variants thereof exactly is involved, typically much more numerically demanding and  available only in few simulations frameworks (see refs.~\cite{parry75a,arnold02a,hu14a,marin-lafleche20a,ahrens-iwers22b} for examples of different implementations).
Therefore, 3d-pbc are often employed and corrected for the interactions across the non-periodic direction either on the fly in the simulations or a posteriori in the calculation of the dielectric profiles.
The former can be achieved using slab corrections, such as the one by \citeauthor{yeh99a}~\cite{yeh99a} or the more general electrostatic layer correction~\cite{arnold02a}.
Alternatively, finite-field approaches \cite{zhang16b} can be employed, which correct for interactions by imposing an additional electrostatic field in order to suppress the electric field over the periodic boundary.
In all of these cases, the linear response equations, \cref{eq:par_fluct_diss,eq:perp_fluct_diss} or alternatively the \enquote{the direct way} via \cref{eq:lin_res_par,eq:lin_res_perp} can be applied straightforwardly.

However, if simulations are performed instead with 3d-pbc, one has to modify the above equations.
Importantly, this is not necessarily an artifact of unwanted interactions across the third dimension, as there are situations where an actually 3d periodic planar system is the appropriate choice, such as stacks of lipid membranes~\cite{schlaich16a,loche20a,wolde-kidan21a}.
In this case the additional field due to the system's dipole has to be subtracted~\cite{stern03a,schlaich16a}.
This results---for the simplest case of tin-foil boundary conditions---in
\begin{equation}
    \label{eq:perp_fluct_diss_3d}
    \epsilon_\perp^{-1} (z) = 1 - \frac{\expval{m_\perp(z) M_\perp} - \expval{m_\perp(z)} \expval{M_\perp}}{\epsilon_0 k_{\text{B}}T + C_\perp/V},
\end{equation}
where $V$ is the system volume and $C_\perp := \expval{M_\perp^2} - \expval{M_\perp}^2$.
The full derivation of this equation is given in section III of the supplementary information~\cite{suppa},
however, it is instructive to mention here that \cref{eq:perp_fluct_diss_3d} follows directly from the effective displacement field being given by $D _{\perp}^{\text{eff}} = D _{\perp}^{\text{ext}} - M_\perp /V$, i.e.\ an external field is weakened by interactions of the system dipole with itself across the $z$-boundary.
Correspondingly, this effective displacement field has to be used in \cref{eq:lin_res_perp} instead of $D_\perp$ in simulations with explicitly applied fields \cite{loche19a}.

Generalizing the tin-foil case above to allow $\epsilon_\infty$ to be of arbitrary value, one finds~\cite{stern03a}
\begin{equation}
    \label{eq:general_perp_fluct_diss}
    \epsilon_\perp^{-1} = 1 - \frac{\expval{m_\perp(z) M_\perp} - \expval{m_\perp(z)} \expval{M_\perp}}{\epsilon_0 k_{\text{B}}T + (\gamma + 1) C_\perp/V},
\end{equation}
where the correction due to the dielectric medium at infinity $\epsilon_\infty$ is given by the correction factor $\gamma = -1/(2\epsilon_\infty + 1) $,
which reduces to \cref{eq:perp_fluct_diss_3d} for $\epsilon_\infty \to \infty$.
Again note that sometimes vacuum boundary conditions, i.e.\ $\epsilon_\infty = 1$, or other values might be used,
whereas common molecular simulation packages such as e.g.\ LAMMPS~\cite{thompson22a},  GROMACS~\cite{abraham15a}, or ESPResSo~\cite{weik19a} select tin-foil boundary conditions by default.

Importantly, \cref{eq:perp_fluct_diss_3d} reduces to \cref{eq:perp_fluct_diss} as $V \to \infty$.
This motivates the use of an additional vacuum layer in $z$-direction in a 3d-pbc simulation without slab corrections and then to correct for interactions across the periodic boundary \textit{a posteriori} through use of \cref{eq:perp_fluct_diss_3d}.
We investigate the influence of the thickness of this additional vacuum layer systematically in the following section, however we note in this context that it is important to actually take $V$ to be the entire system volume---including the vacuum layer---which has not always been done consistently in some works~\cite{jalali20a,ahmadabadi21a,jalali21a}.

\subsection{Investigation of the Impact of Boundary Conditions and Periodicity}
\label{sec:pbc_investigation}

\begin{figure}[h]
    \centering
    \includegraphics[width=\linewidth]{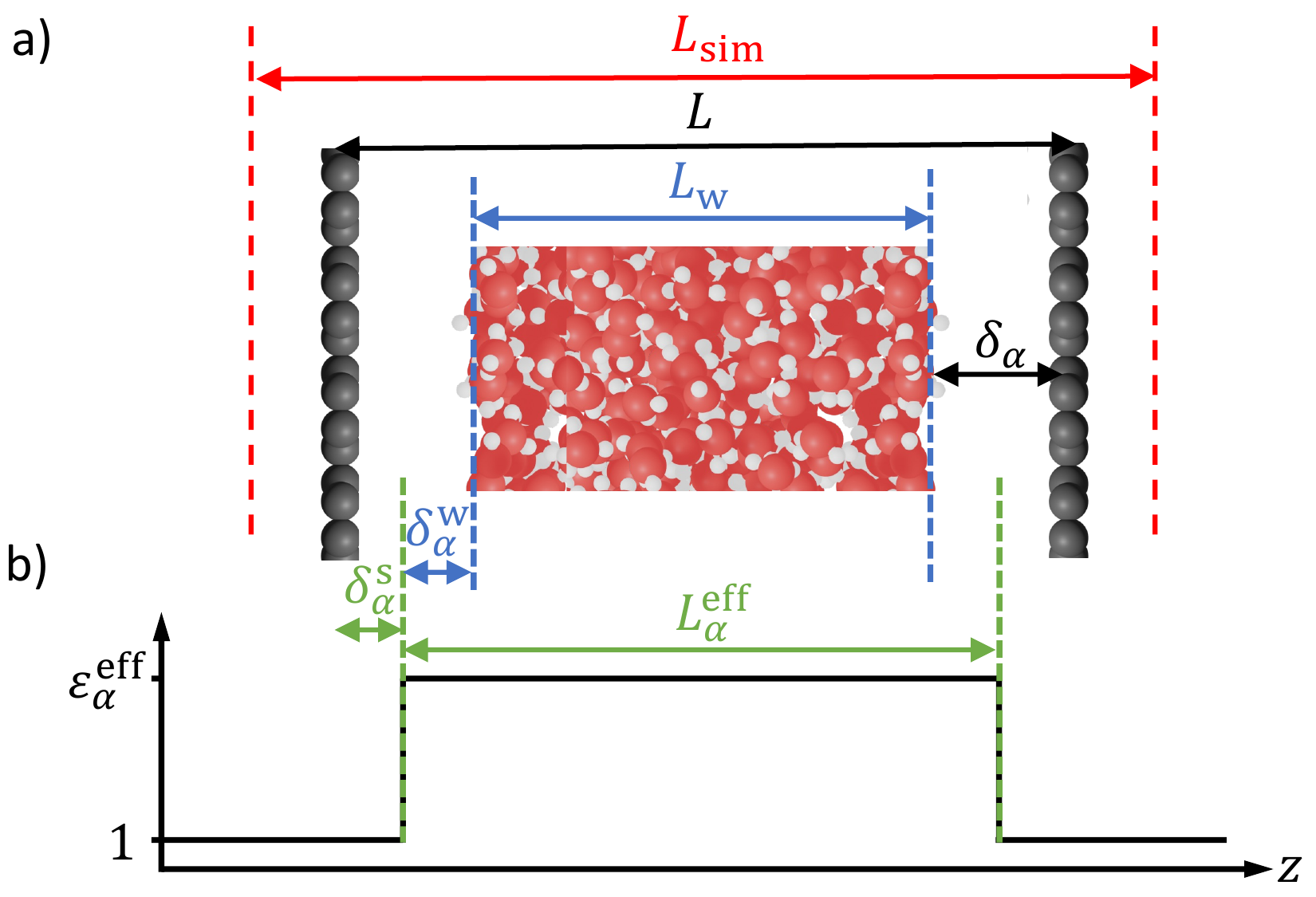}
    \caption{
        Schematic illustration of the different lengths involved and the step profile approach.
        (a) The simulation box size $L_\mathrm{sim}$ can in general be larger
        than the surface atom separation $L$, which is typically taken also as
        plate distance when describing a capacitor.
        The water slab thickness $L_\mathrm{w}$ follows from the Gibbs dividing
        surface, cf.\ \cref{eq:gibbs_length}.
        (b) For coarse-graining a step profile of effective dielectric thickness
        $L_\alpha^\mathrm{eff}$ and effective dielectric constant
        $\varepsilon_\alpha^\mathrm{eff}$ is employed.
        The depletion layer thickness $\delta$ can be decomposed into
        contributions from dielectric interfacial shift $\delta_\alpha^\mathrm{w}$
        and from the Stern layer thickness $\delta_\alpha^\mathrm{S}$,
        as explained in the text and \cref{tab:shifts}.
    }
    \label{fig:dielectric_lengths}
\end{figure}

\begin{figure}[h]
    \centering
    \includegraphics[width=\linewidth]{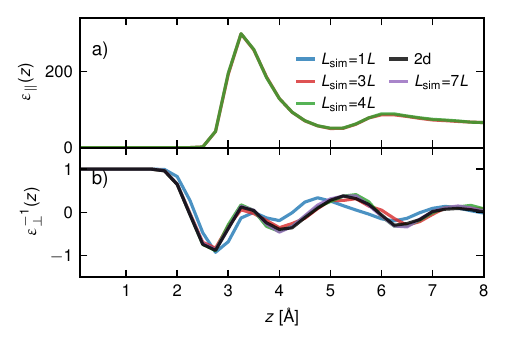}
    \caption{Dielectric profiles at the graphene/water interface for different treatments of the boundary conditions. 
    Labels denote the effective simulation box size $L_\text{sim} = x \cdot L$ for simulation systems that where calculated with 3d periodic boundary conditions. 
    The data labeled with \enquote{2d} is obtained from two-dimensional boundary conditions.
    Data are shown for
        (a) the parallel dielectric profile $\epsilon_\parallel (z)$ (curves overlap)
    and 
        (b) the perpendicular inverse dielectric profile $\epsilon_\perp^{-1}(z)$.
    }
    \label{fig:dielectric_profile_gaps}
\end{figure}

While the expressions for dielectric profiles in different periodicity have been derived and used in previous works, the convergence of 3d-pbc towards the 2d-pbc limit, i.e.\ the sensitivity of the profiles to the width of the vacuum layer, has not yet been systematically investigated.
To address this, here we simulate 1871 SPC/E water molecules confined between graphene sheets separated by $L=\SI{34}{\angstrom}$, see \cref{fig:dielectric_lengths}(a) and the Supporting Information~\cite{suppa} for details.
The results of these simulations are summarized in \cref{fig:dielectric_profile_gaps}.
Since the periodic interactions only affect the perpendicular dielectric profiles, the parallel components shown in \cref{fig:dielectric_profile_gaps}(a) perfectly agree in all cases.
However, the inverse perpendicular profiles in \cref{fig:dielectric_profile_gaps}(b) show subtle differences with the length of the vacuum slab.
Without an additional vacuum layer, $L_\text{sim}=L$ [blue data labeled in \cref{fig:dielectric_profile_gaps}(b)], this setup corresponds to an infinite stack of water slabs directly adjacent to each other, separated only by a double graphene layer.
This is markedly different from a two-dimensional periodic system because the water slabs interact with each other via both electrostatic and Lennard-Jones interactions.

According to \cref{eq:perp_fluct_diss_3d} the 3d-pbc profiles should converge to that obtained with 2d-pbc if $L$ is sufficiently large.
Indeed, the inverse perpendicular profile for $L_\text{sim} = 3 \cdot L$ still deviates slightly from those of the factor three and six. 
However, for $L _{\text{sim}} \ge 4 \cdot L$ the deviations from the two-dimensional result are practically negligible.
This justifies the use of 3d-pbc simulations to calculate dielectric profiles of 2d-pbc systems, provided the vacuum length is chosen suitably.
The above considerations apply for both approaches, i.e.\ utilizing equilibrium fluctuations (see e.g.\ refs.~\cite{stern03a,majumdar21a,schlaich16a,loche20a}) or the calculation of dielectric profiles the \enquote{direct way}~\cite{loche19a}.
%

\section{A brief Historical Perspective}
While the connection between molecular details and dielectric behavior for bulk fluids has been subject to research for decades \cite{Nienhuis71,Pollock81,bopp96a,Bopp98}, the detailed microscopic investigation of interfacial and confinement effects only evolved around the turn of this century.
Starting with the seminal work of \citeauthor{stern03a}~\cite{stern03a}, the concept of static dielectric permittivity profiles were introduced.
At this time, experimental results already highlighted anomalous dielectric behavior from atomic force microscopy measurements~\cite{teschke01a}.
Following work utilized the equilibrium fluctuation relations in order to calculate the parallel dielectric profiles for various systems~\cite{ballenegger05a,froltsov07a} or measured the polarization response of fluids the \enquote{direct-way}~\cite{faraudo04a}.
The determination of $\epsilon_\perp^{-1}(z)$ seems to have been out of reach for the computational power at that time due to insufficient statistics available~\cite{ballenegger05a}.
The first report on both components of planar dielectric profiles was given by \citeauthor{bonthuis12a}~\cite{bonthuis12a}, who also showed that dielectric profiles can rigorously be derived without the need of locality assumptions (see \cref{sec:determination_profiles}).

Thereafter, interest in the field grew considerably and over the next ten years numerous publications considered dielectric profiles for various systems and various force-fields and water models~\cite{gekle12a,ghoufi12a,zhu12a,gekle13a,parez14a,renou15a,schlaich16a,meneses-juarez18a,sato18a,loche19a,loche19b, vargheseEffectHydrogenBonds2019, zhu20a,loche20a,motevaselian20b,jalali20a,monet21a,ahmadabadi21a,majumdar21a,hu21a,mondalAnomalousDielectricResponse2021, jalali21a,zhu22a,mulpuri23a,borgisDielectricResponseConfined2023,dinpajooh23a,becker24a, tang_scale-dependent_2025}.
In \cref{tab:water_models}, we provide an overview of the different water models that have been used in the context of dielectric profiles.
Authors used different methods to elucidate the dielectric response of nano-confined fluids via free energy considerations~\cite{nymeyer08a,cox22a} or by determining an effective dielectric constant directly, without considering microscopically resolved profiles~\cite{zhang13a,itoh15a,deluca16a,mondal19a,dinpajooh23a}.
Such averaging procedures have to be interpreted carefully, as we discuss in detail in \cref{sec:coarse_graining}.



However, the extensive amount of research has introduced some confusion regarding the appropriate use of boundary conditions and the correct fluctuation relations for specific simulation setups, which we have introduced in detail above.
Some studies have adjusted the volume term for the correction of 3d-pbc~\cite{jalali20a,ahmadabadi21a,jalali21a}, while others modified the fluctuation-dissipation relation for the boundary condition employed in their simulations~\cite{guo21a,mondal19a,zhu22a,deluca16a}. 
Additionally, certain reports have interchanged the fluctuation relations for 3d- and 2d-pbc~\cite{majumdar21a}, or assumped a divergence-free displacement field in systems with free charges~\cite{hu21a,renou15a,jalali21a,zhu22a}.

Dielectric permittivity profiles have also been investigated for other fluids, including methanol and dichloromethane~\cite{motevaselian20b}, as well as Stockmayer fluids~\cite{froltsov07a,borgisDielectricResponseConfined2023}.
Some studies have examined dielectric profiles of water confined by metallic walls with an applied potential~\cite{deissenbeck21a} and the concept of dielectric profiles has also been extended to different geometries, such as cylindrical~\cite{zhu12a,gekle13a,loche19b,ghoufi12a,zhu22a,zhu20a,renou15a} and spherical confinement~\cite{ballenegger05a,schaaf15a,mondal19a,ghoufi12a}.
Further research has been performed on the dielectric behavior of fluids surrounding spherical solutes~\cite{gekle12a,dinpajooh23a}, including frequency and location-dependent dielectric properties~\cite{gekle12a}.
Recent studies have started exploring the dielectric properties of nanoconfined water using ab-initio simulations~\cite{ruiz-barragan20a,deissenbeck23b} and machine-learned force-field approaches~\cite{dufils24a}.

Sustainable software workflows that allow for following the FAIR data principles \cite{wilkinson_fair_2016} have successfully established  the extraction of permittivity profiles from molecular simulations in different geometries \cite{maicos}.
However, so far no consensus seems to have been reached on the interpretation of the simulation results and their connection to experimental observables, as we discuss in the following.

\section{Coarse-Graining Profiles Through Effective Medium Theory and Equivalent Circuits}
\label{sec:coarse_graining}

\subsection{Step profile approach}
\label{sec:step_profile_approach}

In order to investigate the consequences for the resulting electrostatic interactions, coarse-grained modeling or experimentally accessible quantities like the capacitance, the complex profiles $\epsilon_\parallel(z)$ and $\epsilon_\perp^{-1}(z)$ typically need to be replaced by analytically tractable expressions.
This is usually done in terms of step profiles $\epsilon_\alpha^\star (z)$ as shown in \cref{fig:dielectric_lengths}(b) with $\alpha=\parallel, \perp$,
complemented by using effective medium theory~\cite{schlaich16a,dufils24a,cox22a,borgisDielectricResponseConfined2023,monet21a,loche20a,becker24a,ahmadabadi21a,loche19b,deissenbeck21a,zhang18a, tangScaledependentAnomalousBehavior2024}.
The corresponding anisotropic tensorial Green's function can explicitly be solved for~\cite{loche20a,kondrat10a,goduljan14a,mohammadzadeh15a}.

\hlork{
Such a procedure is further motivated by the fact that local non-linear response variations within the interface, 
which take place at Angstrom scale, are experimentally accessible with depth-resolved sum-frequency-generation spectroscopy \cite{fellowsHowThickAir2024},
whereas the  local linear response cannot be probed directly.
However, the integral over the perpendicular $E$ field, that is the integral over the inverse perpendicular dielectric function, gives the surface potential \cite{kornyshevEffectSpatialDispersion1981}, which is experimentally measurable.
That is, only the integral over the local dielectric profiles needs to be reproduced by any step profile approach to reproduce the many observables.
}

Since all of such step function approaches suffer from ambiguous definitions of the interface position~\cite{dufils24a,cox22a}, it is tempting to replace the complex profile by a dielectric medium with effective dielectric constants $\epsilon_\alpha^\text{eff}$ and effective widths $L_\alpha^\text{eff}$, such that 
\begin{equation}
    \epsilon_\alpha^\star(z) = \begin{cases}
        \epsilon_\alpha^\text{eff} & \text{for } |z| < L_\alpha^\text{eff}/2, \\
        1 & \text{for } |z| > L_\alpha^\text{eff}/2.
    \end{cases}
\end{equation}
To derive the corresponding values $\epsilon_\alpha^\text{eff}$ and $L_\alpha^\text{eff}$, concepts of effective medium theory can be employed as follows~\cite{hasted73a}.
In a thermodynamic sense, this essentially corresponds the concept of Gibbs' adsorption theory to construct a dividing surface~\cite{gibbs61a}.
The basic idea is that a coarse-grained slab with these effective parameters should exhibit the same response as the nanoconfined fluid when averaged over the slab~\cite{schlaich16a}.
Since in the parallel direction the electric field is constant, cf. \cref{eq:par_eps_start}, the integral over the displacement $D_\parallel(z)$ must be reproduced, yielding
\begin{equation}
    \label{eq:parallel_macroscopic_profile}
    \epsilon_{\parallel}^{\text{eff}} = 1 + \frac{ \int_{-L/2}^{L/2} \epsilon_\parallel (z) \dd z - L }{ L_{\parallel}^{\text{eff}} },
\end{equation}
where $L_{\parallel}^{\text{eff}}$ is the effective length of the slab model, $\epsilon_{\parallel}^{\text{eff}}$ is the effective dielectric constant in parallel direction of the slab and $L$ is the distance in $z$-direction between the positions where the response is measured.
Similarly, for the perpendicular case the displacement field is constant, cf.\ \cref{eq:perp_eps_def}.
Thus, the electrostatic potential drop across the system---or, correspondingly, the system's capacitance---in $z$-direction must be reproduced,
\begin{equation}
    \label{eq:perpendicular_macroscopic_profile}
    \left( \epsilon_{\perp}^{\text{eff}}\right)^{-1} = 1 + \frac{ \int_{-L/2}^{L/2} \epsilon_\perp^{-1} (z) \dd z - L }{ L_{\perp}^{\text{eff}} },
\end{equation}
where we have introduced the effective length $L_{\perp}^{\text{eff}}$ and the effective dielectric constant $\epsilon_{\perp}^{\text{eff}}$ in the perpendicular direction.
Importantly, \cref{eq:perpendicular_macroscopic_profile} reveals that it is not sufficient to simply integrate over the profile $\varepsilon_\perp^{-1}(z)$ to obtain an averaged dielectric permittivity \cite{mondalAnomalousDielectricResponse2021, choganiEffectSalinityDielectric2024, coelho_interplay_2024, fuentes-azcatl_dielectric_2025}, since this does not reproduce the potential drop.

A few remarks are in order here:
First, one is often interested in the perpendicular macroscopic response as this is the quantity of interest for capacitors and can be compared to experimental measurements, as for instance those by \citeauthor{fumagalli18a}~\cite{fumagalli18a}.
Noteworthy, recent experiments by \citeauthor{wang24b} have also measured the parallel dielectric response~\cite{wang24b}.
Second, there are two unknown variables in each of the above relations, the effective lengths $L_{\alpha}^{\text{{eff}}}$ and the effective dielectric constants $\epsilon_{\alpha}^{\text{{eff}}}$, so \cref{eq:parallel_macroscopic_profile,eq:perpendicular_macroscopic_profile} are under-determined without additional assumptions.
A common choice is to set $L _{\perp}^{\text{eff}} = L$, with the latter referring to the carbon-carbon distance between the innermost atoms of the confining graphene sheets, i.e.\ $\epsilon_{\perp}^{-1} (z)$ is averaged over some effective pore width~\cite{deluca16a}.
Yet, such a choice for the system size is not rigorous and other definitions are possible~\cite{itoh15a,zhang13a}.
Importantly, the effective dielectric constant $\epsilon_{\alpha}^{\text{eff}}$ resulting from such a procedure does not reflect the local values inside the slab.
The averaging procedure can also be refined by partitioning the system into interfacial and bulk layers, corresponding to an equivalent circuit of serial capacitors~\cite{dufils24a}.

\begin{figure*}[htbp]
    \centering
    \includegraphics[width=.49\textwidth]{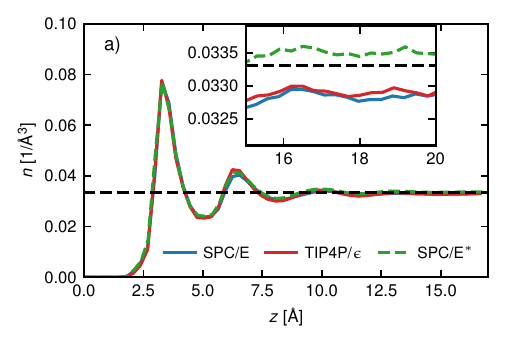}
    \includegraphics[width=.49\textwidth]{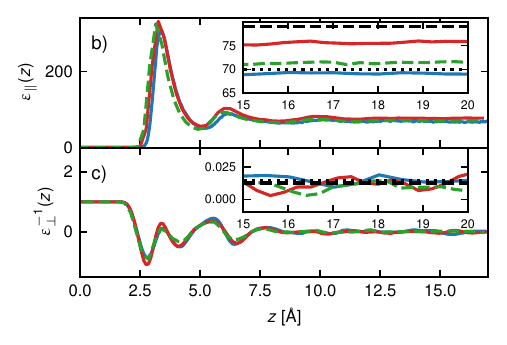}
    \caption{(a): Number density profiles of the TIP4P/$\epsilon$ and SPC/E water models in a system with $L = \SI{34}{\angstrom}$. The SPC/E* results were obtained from simulations with a higher number density (see below) compared to SPC/E and TIP4P/$\epsilon$, which were simulated at the same density.
    Black line denotes the experimental water density of $n = \SI{0.033}{\angstrom^{-3}}$~\cite{wagner02a}. Due to the system's symmetry, only one half of the simulation box is shown.
    Corresponding parallel and inverse perpendicular dielectric profiles are shown in (b) and (c), respectively.
    The insets show data for $L=60\,\text{\AA}$ revealing constant profiles that differ slightly from the respective bulk values (dashed line in (a) and values in \cref{tab:water_models}, in (b) and (c), the dashed line corresponds to the TIP4P/$\epsilon$ bulk value, the dotted to the SPC/E bulk value of the dielectric constant.)
    }
    \label{fig:density_plot}
\end{figure*}

The choice of the effective length $L_{\alpha}^\text{eff}$ sensitively affects the effective macroscopic response of the nanoconfined fluid---or, vice versa, is a free parameter to deduce $\epsilon_{\alpha}^\text{eff}$, both in simulations and in experiments~\cite{loche20a}.
This can be circumvented by employing concepts of surface thermodynamics and effective medium theory as follows~\cite{schlaich16a}:.
Since the fluid in the pore center behaves bulk-like (cf.\ \cref{fig:example_epsilon_profs,fig:density_plot}), the relative contribution of the interfacial region diminishes as the pore size increases.
Consequently, in the limit of large slabs, $L \to \infty$, the effective dielectric constant of the confined fluid must recover its bulk value, $\epsilon_{\alpha}^{\text{eff}} \to \epsilon_{\text{bulk}}$, thereby uniquely fixing $L_{\alpha}^{\text{eff}}$.
We then use this effective length to calculate the dielectric interfacial shift $\delta_\alpha^\text{w}$ according to \cite{bonthuis11a}
\begin{equation}
    \label{eq:correction_length}
    2\delta_\alpha^\text{w} := L_{\alpha}^\text{eff} - L_{\text{w}},
\end{equation}
where the water slab thickness is defined based on the Gibbs-dividing-surface distance~\cite{gibbs61a}
\begin{equation}
    \label{eq:gibbs_length}
L_{\text{w}} := \frac{N_{\text{w}}}{{A n_{\text{bulk}}}} ,
\end{equation}
with $N_{\text{w}}$ the number of water molecules in the slab, $A=1885.95\,\text{\AA}^2$ its area in lateral directions and $n_{\text{{bulk}}} = 0.03336\,\text{\AA}^{-3}$ the bulk number density.
The corresponding depletion layer thickness $\delta_{\alpha}$, shown in \cref{fig:dielectric_lengths}, follows rigorously from a Gibbs construction for two bulk phases that meet at an interface~\cite{janecek07a}.
In practice, $\delta_\alpha^\text{w}$ converges to a constant value quickly with increasing pore widths $L$ and for the system considered here we find $L=\SI{60}{\angstrom}$ sufficiently large in line with our previous studies~\cite{loche20a}, revealing that at these separations upon increasing the pore size the additional water in the slab is bulk-like.

\begin{figure*}[htbp]
    \centering
    \includegraphics[width=.49\textwidth]{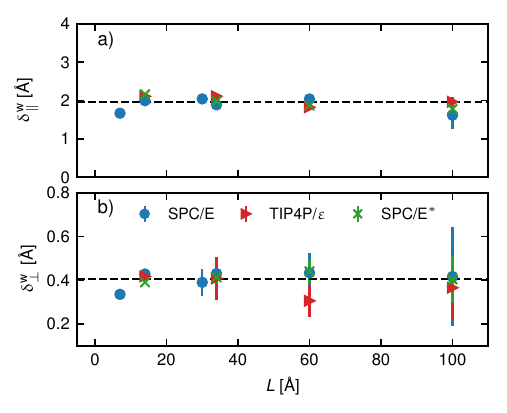}
    \includegraphics[width=.49\textwidth]{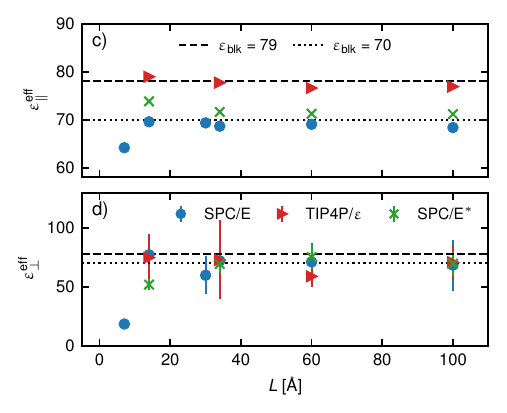}
    \caption{Panels a) and b): Interfacial dielectric shifts $\delta_\alpha^\text{w}$ for the different water models and densities in the graphene slit pore.
    Dashed lines are obtained by averaging all data for $L>10\,\text{\AA}$, yielding $\delta_\parallel^\text{w} = 1.92 \pm 0.09\,\text{\AA}$ and $\delta_\perp^\text{w} = 0.40 \pm 0.07\,\text{\AA}$, cf.\,\cref{tab:water_models}.
    Panels c) and d): Parallel and perpendicular effective dielectric constant of different water models as a function of graphene plate distances $L$.
    Error bars are obtained from rigorous error propagation of the statistical uncertainty.
    The dashed and dotted lines denote the corresponding values of the water models in bulk.
}
    \label{fig:epsilon_eff_summary}
\end{figure*}

Exemplary density and dielectric profiles for $L=\SI{34}{\angstrom}$ are shown in \cref{fig:density_plot} for the TIP4P/$\epsilon$ and the SPC/E water models.
Importantly, the dielectric constant of water depends sensitively on its density \cite{uematsuStaticDielectricConstant1980, florianoDielectricConstantDensity2004, marshallDielectricConstantWater2008}, which in confinement is generally challenging to control, as will be elaborated further below.
Indeed, the density in the bulk-like region between the graphene sheets at $z>15\text{\AA}$ differs slightly from the corresponding values in bulk simulations [inset of \cref{fig:density_plot}(a)].
We therefore average the parallel dielectric permittivity (since averaging $\varepsilon_\perp^{-1}(z)$ is too noisy) in the bulk-like region at least $15\,\text{\AA}$ away from the interface for $L\geq 60\,\text{\AA}$, where the profiles are constant, in order to estimate the values in agreement with the in-pore water density.
Such averaging yields $\varepsilon = 68.77 \pm 0.14$ and $\varepsilon = 76.50 \pm 0.12$ for SPC/E and TIP4P/$\epsilon$, respectively, in reasonable agreement with the bulk value (\cref{tab:water_models});
the value $\varepsilon = 71.36 \pm 0.02$ for SPC/E$^\star$ corresponding to a slightly higher water density will be discussed below.

By taking into account the bulk dielectric constant corresponding to the in-pore water density, we can now employ \cref{eq:parallel_macroscopic_profile,eq:perpendicular_macroscopic_profile} and demand that the box profile reproduces these corresponding values, thereby obtain the interfacial shifts $\delta_\alpha^\text{w}$ according to \cref{eq:correction_length}, as is shown in \cref{fig:epsilon_eff_summary}(a).
Surprisingly, all data converge for $L>10\,\mathrm{\AA}$, yielding $\delta_\parallel^\text{w} = 1.92 \pm 0.09\,\text{\AA}$ and $\delta_\perp^\text{w} = 0.40 \pm 0.07\,\text{\AA}$, as reported in \cref{tab:shifts_materials}.
This reveals, that interfacial dielectric effects are, as already observed from the profiles shown in \cref{fig:density_plot}, rather insensitive to the water model employed, provided that the bulk dielectric constant is taken into account correctly.

\begin{table}[htbp]
\caption{Water models employed to study the static dielectric response of confined water. The middle column lists approximate values for the bulk dielectric constant $\epsilon _{\text{bulk}}$ of the water models at $T = \SI{300}{\kelvin}$.}
\label{tab:water_models}
\begin{tabular}{@{}lcl@{}}
\toprule
Water model & $\approx \epsilon _{\text{bulk}}$ & Confinement effects studied \\ \midrule
SPC/E       & $70.0\pm 0.2$\cite{schlaich16a,ramireddyDielectricConstantSPC1989, gerebenAccurateCalculationDielectric2011, kusalik94a} & \begin{tabular}{@{}l@{}}\cite{loche18a,loche19a,loche20a,schlaich16a,bonthuis12a,cox22a,cox20a}\\ \cite{zhang18a,deluca16a,motevaselian20b,majumdar21a,parez14a,jalali20a, borgisDielectricResponseConfined2023}\end{tabular} \\
SPC/$\epsilon$ & 80~\cite{fuentes-azcatl15a}   &  \cite{meneses-juarez18a}         \\ 
TIP3P       & 97 \cite{braun_transport_2014,izadiBuildingWaterModels2014,liese_dielectric_2022} &  \cite{deissenbeck21a}         \\
TIP4P/2005  & 60~\cite{abascal05a} &   \cite{bonthuis12a,hu21a,renou15a,majumdar21a,majumdar21a,parez14a}        \\
TIP4P/Ew    & 65~\cite{horn04a} &  \cite{loche18a,renou15a}         \\ 
TIP4Q    & 80~\cite{alejandre11a} &  \cite{meneses-juarez18a}         \\ 
TIP4P/$\epsilon$ & $79.3\pm 0.23$~\cite{fuentes-azcatl14a}   & This work         \\  \bottomrule
\end{tabular}
\end{table}

\subsection{From interfacial dielectric shifts to the Stern layer thickness and effective dielectric constants}

\begin{table}[h]
\caption{Dielectric shifts for different interfaces. 
The values for graphene are determined by averaging the data shown in \cref{fig:epsilon_eff_summary}a,b for $L>10\,\text{\AA}$.}
\centering
\begin{tabular}{@{}lllll@{}}
\toprule
{System} & Ref. & $ \delta^{\text{w}}_\perp$ [$\si{\angstrom}$] & $\delta^{\text{w}}_\parallel$ [$\si{\angstrom}$] \\
\midrule
Graphene & This work & $0.40 \pm 0.07$    & $1.92 \pm 0.09$ \\
Diamond (hydrophilic)\footnotemark[1] & \cite{bonthuis12a}          &-0.3   & -0.2 \\
Diamond (hydrophobic)\footnotemark[1] & \cite{bonthuis12a}         & 1     & 1.4  \\
Vapor                                 & \cite{loche22b}             & 2.75          & 0.5                      \\
Decanol\footnote[1]{water contribution without polarization of the surface groups}  & \cite{schlaich16a}    & 0.075    & 0.6   \\
DMPC\footnotemark[1]                                                                & \cite{loche20a}       & -2.5     & -0.5  \\
DGDG\footnotemark[1]                                                                & \cite{loche20a}       & 3        & -0.4   \\
Uncharged silica\footnotemark[1]      & \cite{hunger22a}           & 0.8   & --                      \\
\cdashline{1-4} 
Hexane                                & \cite{loche22b}             & 3.4~          & 2.2                      \\
Decanol\footnote[2]{including surface polarization}                                 & \cite{schlaich16a}    & 3.5      & 2  \\
DMPC\footnotemark[2]                                                                & \cite{loche20a}       & 7        & 23    \\
DGDG\footnotemark[2]                                                                & \cite{loche20a}       & 9.5      & 0.75    \\
Uncharged silica\footnotemark[2]      & \cite{hunger22a}            & 1.4           & --                      \\
\bottomrule
\end{tabular}
\label{tab:shifts_materials}
\end{table}

\Cref{tab:shifts_materials} summarizes the dielectric interfacial shifts $\delta_\alpha^\text{w}$ for different aqueous interfaces.
For graphene, both $\delta_\perp^\text{w}$ and $\delta_\parallel^\text{w}$ are positive, revealing that the confined system screens electrostatic interactions more efficiently than a bulk-like system of the same slab thickness.
In contrast, for example the hydrophilic diamond surface shows a negative shift, related to strong water layering at the interface and thus to a water slab thickness $L_\mathrm{w} > L_\alpha^\mathrm{eff}$, i.e.\ the dielectric screening of a bulk-like slab of width $L_\mathrm{w}$ is more efficient than that of the confined water.

Another observation that can be drawn from the data shown in \cref{tab:shifts_materials} is that at the soft water-vapor interface $\delta_\parallel^\mathrm{w} > \delta_\perp^\mathrm{w}$, which can be related to the characteristic water orientation at such interfaces \cite{becker24a}.
Importantly, if the aqueous interface is with another dielectric (e.g.\ with hexane or a lipid bilayer), the polarization of the surface groups needs to be included in the calculation of the dielectric interfacial shifts to recover the system's effective response.
Nevertheless, it is instructive to separate the water contribution from the surface polarization, as this allows for a direct comparison of the dielectric interfacial shifts with those of other aqueous interfaces in \cref{tab:shifts_materials}.
At highly polar interfaces such as at the phosphocholine lipid DMPC, the water contribution to the dielectric excess is negative, i.e.\ the dielectric response is reduced with respect to bulk water.
However, if the polarization due to the lipids is included, the dielectric shifts are large, with $\delta_\perp^\text{w} = 7$~\AA{} and $\delta_\parallel^\text{w} = 23$~\AA{}, revealing strong in-plane dipole fluctuations of the lipids.
This is different for the digalactosyldiacylglycerol lipid DGDG, which shows a small in-plane shift $\delta_\parallel^\text{w} = 0.75$~\AA{} due to the absence of a strong dipole moment, but where the water perpendicular shift $\delta_\perp^\text{w} = 3$~\AA{} is significant due to strong hydrogen bonding \cite{kanduc17a}.

For the calculation of the capacitance, which we address below, one is however rather interested in the shift between the surface separation $L$ (defining a classical plate capacitor) and the dielectric effective thickness $L_\perp^\text{eff}$ than in the width of the depletion layer $\delta_{\alpha}$, i.e.\ in the Stern layer thickness $\delta_\perp^\text{S}$~\cite{stern24a}.
The definition of the shifts shown in \cref{fig:dielectric_lengths} is summarized in \cref{tab:shifts}.
The Stern layer essentially characterizes the difference between the region where the dielectric response of the fluid is probed and the definition of the pore width $L$, in line with Stern's original idea of a dielectrically dead layer.

\begin{table}[h]
    \caption{Relation of the dielectric interfacial shift $\delta_\alpha^\text{w}$ to the Stern layer thickness $\delta_\alpha^{s}$ and the depletion layer thickness $\delta$.}
    \label{tab:shifts}
    \begin{tabular}{@{}lcl@{}}
    \toprule
    $2\delta_\alpha^\text{w} = L_\alpha^\text{eff} - L_\mathrm{w}$ & dielectric interfacial shift \cite{bonthuis11a} \\ 
    $2\delta_\alpha^\text{s} = L - L_\alpha^\text{eff}$ & Stern layer thickness \\ 
    $\delta = \delta_\alpha^\text{s} + \delta_\alpha^\text{w}$ & depletion layer thickness \\
      \bottomrule
    \end{tabular}
    \end{table}

For water at graphene interfaces, we find the Stern layer thickness $\delta_\perp^\text{S} = \SI{1.50}{\angstrom}$ to be the dominant contribution to the depletion layer thickness $\delta_\perp = 1.90\,\text{\AA}$, whereas $\delta_\perp^\text{w} = \SI{0.40}{\angstrom}$, i.e.\ the perpendicular dielectric dividing surface position is close to the Gibbs dividing surface in \cref{fig:example_epsilon_profs}.
As we will discuss below, this Stern layer thickness has a dramatic impact on the resulting capacitance of the aqueous graphene capacitor.
The parallel Stern layer thickness $\delta_\parallel^\text{S} = \SI{-0.06}{\angstrom}$ is negligibly small and thus has only minor impact on the in-plane capacitance \cite{wang24b}.
The fact that it is negative, however, reflects that the effective dielectric thickness $L_\parallel^\text{eff}$ is larger than the pore width $L$, i.e.\ the position of the parallel dielectric dividing surface is at $z<0$ in \cref{fig:example_epsilon_profs}(c).
Consequently, the parallel depletion layer thickness $\delta_\parallel = 1.98\,\text{\AA}$ is of similar magnitude as the perpendicular one, but is dominated by the dielectric interfacial shift $\delta_\parallel^\text{w} = \SI{1.92}{\angstrom}$.

In the next step, having assessed the dielectric interfacial shifts that are constant for sufficiently large pore sizes ($\gtrsim 1\,\mathrm{nm}$ for graphene!), the effective dielectric thickness $L_{\alpha}^\text{eff}$ for all pore sizes is known and allows to determine the corresponding values $\epsilon_{\alpha}^\text{eff}$ via \cref{eq:parallel_macroscopic_profile,eq:perpendicular_macroscopic_profile}, shown in \cref{fig:epsilon_eff_summary}(b).
Noteworthy, in line with the constant dielectric interfacial shifts, the effective dielectric response of water in planar confinement is constant and bulk-like down to pore widths as small as about 1$-2$\,nm depending on the surface chemistry \cite{schlaich16a,loche20a}.
This reveals, that even for separations where the layering in the density and dielectric profiles for water at graphene interfaces is overlapping (\cref{fig:density_plot}), the slab-averaged dielectric response can be expressed in terms of the water bulk behavior, provided that the slab width is chosen consistently.
The interfacial portions and full profiles for all pore sizes are shown in the SI in section I~\cite{suppa}.

\subsection{Impact of the Water Model and Quantity on the Dielectric Response}
\label{sec:water_quantity}

We now turn to a careful comparison of the influence of the simulation details on the resulting separation-dependent dielectric response of water confined between the inert graphene sheets.
Whereas the majority of simulations have utilized the classical SPC/E water model,
a wide variety of other water models have been employed, with significant spread in the bulk water dielectric behavior, see~\cref{tab:water_models}.
Also machine-learned force fields from ab-initio data have been employed recently to study the water dielectric response between planar graphene sheets~\cite{dufils24a}.
The choice of the water model and the detailed simulation parameters not only affects static properties like density or dielectric response, but also has large impact on the resulting thermodynamics and dynamic properties, which is well discussed in the literature~\cite{tsimpanogiannisSelfdiffusionCoefficientBulk2019, leeTransportPropertiesBulk2019, desgrangesBenchmarkFreeEnergies2017, taziDiffusionCoefficientShear2012, fanourgakisDeterminingBulkViscosity2012}.
Notably, to the best of our knowledge, the TIP4P/$\epsilon$ water model has not been investigated in the context of dielectric response in planar confinement, which seems surprising given that it was optimized to match the experimental bulk dielectric behavior of water.
Independent simulations performed for the SPC/E model at 300~K yield in a bulk dielectric constant of $\epsilon = 70 \pm 0.2$ \cite{schlaich16a}, whereas the TIP4P/$\epsilon$ water model yields $\epsilon = 79.3 \pm 0.2$, close to the experimental value of $ \epsilon= 77.75$~\cite{haynes16a}.
Furthermore, the model has been shown to match important thermodynamic and dynamic experimental observables well, across a large range of temperatures~\cite{fuentes-azcatl14a}.

In \cref{fig:density_plot}(a), we compare the number density profiles of both SPC/E water and TIP4P/$\epsilon$ near a graphene sheet.
Interestingly, they appear to be very similar, except for a slight change in the second layer density peak, where the density of TIP4P/$\epsilon$ is marginally larger than that of SPC/E.
The dielectric profiles of both water models for $L = \SI{34}{\angstrom}$, are shown in \cref{fig:density_plot}(b,c).
The profiles $\varepsilon_\parallel(z)$ in \cref{fig:density_plot}(b) follow the similarity between both water models, with increase in the second peak for TIP4P/$\epsilon$ being proportional to the increased density.
\hlork{
    As shown in the inset of \cref{fig:density_plot}(b), the profiles converge to constant values that are close to the bulk values of the respective water models, as expected.
}
Also $\varepsilon_\perp^{-1}(z)$ in \cref{fig:density_plot}(c) shows consistent positions of the zero crossings and extrema and slightly enhanced peak magnitude for TIP4P/$\epsilon$\hlork{, while being consistent with their respective bulk dielectric behavior away from the interface, which is, however, hard to resolve within the simulations}.
It is thus tempting to compare both water models with regard to the dielectric shifts and effective dielectric behavior in \cref{fig:epsilon_eff_summary} (also see the asymptotic shifts in \cref{tab:shifts_materials}).
The TIP4P/$\epsilon$ water model shows remarkably similar behavior when compared to the results of SPC/E and thus motivates defining a dielectric shift independent of the water model as done above.
We re-iterate the importance for taking the precise bulk behavior into account, reflected by the convergence of the models to their respective bulk values in \cref{fig:epsilon_eff_summary}(c,d).

Crucially, in a confined system the number of water molecules needs to be chosen with care.
Since water is practically incompressible at ambient conditions, small changes in the number of confined molecules alter the system's pressure---and, thus, it's thermodynamic state---drastically.
All results for the SPC/E water model discussed so far were obtained by mimicking the contact with a (fictitious) external bulk reservoir, i.e.\ controlling the chemical potential of water \cite{loche20a}.
The latter can be achieved using techniques from grand-canonical simulations, such as e.g.\ the use of thermodynamic integration \cite{schlaich16a, loche20a} or hybrid GCMC/MD schemes \cite{renou15a}.
Another common approach is to control the pressure on the confining walls, either by varying the particle number or the graphene distance to achieve a prescribed pressure \cite{ruiz-barragan20b}.
Note, that a constant perpendicular pressure corresponds to an experimental setup where pores are flexible, such as membranes that are allowed to swell~\cite{kanduc14a, schlaich15a}.
In typical measurements of water confined between graphene, the pore loading will rather follow from equilibrium of the chemical potential.
Importantly, both approaches correspond to different thermodynamic ensembles and might lead to different amounts of water confined in the pores.
Other studies of confined water's dielectric behavior also employ less physically motivated approaches, such as varying pore size or particle count until the expected bulk density is observed far from the interface~\cite{parez14a}.

The simulations for the TIP4P/$\epsilon$ model discussed above were performed with the same number of water molecules as determined for the SPC/E model in ref.\ \cite{loche20a}.
We now study separately the influence of the water number $N_\mathrm{w}$ on the dielectric properties.
To this end, we performed a second set of simulations using the SPC/E water model with different numbers of water molecules, denoted \enquote{SPC/E$^*$}.
In detail, the water amount was determined from simulations in explicit contact with a bulk reservoir, see Supplementary Information~\cite{suppa}.
While this also should mimic chemical equilibrium at the same conditions, we attribute the difference in $N_\mathrm{w}$ to the fact that due to the periodic reservoir, no slab correction for the electrostatic interactions was employed.
The mild change of the water number has negligible effect on the interfacial water structuring (\cref{fig:density_plot}(a)), yet subtle differences in the dielectric profiles can be observed(\cref{fig:density_plot}(b,c)), related to the different water densities in the bulk-like region.
Interestingly, we observe that our previous data relying on extrapolation of the chemical potential to determine $N_\mathrm{w}$ yields densities about 1.5\% smaller than expected from bulk.
We speculate, that this discrepancy might be an artifact that arises from a slight shift of the chemical potential due to slab correction for the electrostatic interactions as this has not been observed in 3-D periodic simulations \cite{schlaich16a,kanduc17a,kowalikHydrationRepulsionDifference2017}.
However, such a slightly different density is sufficient to reduce the dielectric response of the confined water as discussed in \cref{sec:step_profile_approach}.
The same applies to the gently higher density (less than 0.5\%) obtained from piston simulations, slightly increasing the dielectric response.
The origin for this increase might be in the finite size effects due to the periodicity of the system, however, these effects are out of the scope of this review and we merely want to stress that taking these effects into account is crucial for any interpretation of the dielectric properties.

Noteworthy, as already introduced above, this different dielectric behavior in the bulk-like region of the confined water does not manifest in the dielectric shifts $\delta_\alpha^{\mathrm{w}}$.
In \cref{fig:epsilon_eff_summary}, we summarize the measured effective dielectric constants of both experiments.
The effective dielectric constants $\epsilon_{\parallel}^{\text{eff}}, \epsilon_{\perp}^{\text{eff}}$ show similar behavior upon varying the pore widths for all the water models/water numbers considered.
We thus conclude that the number of particles has no significant influence on the effective dielectric behavior or the shift of the dielectric dividing surface if analyzed carefully with respect to eventual differences in the bulk-like regions.
However, varying the number of particles beyond the range studied here might lead to different interfacial structuring and thus great care should still be taken to match the desired thermodynamic conditions with appropriate particle numbers in simulations of confined fluids.

\subsection{Sensitivity on the Effective Pore Size}
\begin{figure}
    \centering
    \includegraphics[width=\linewidth]{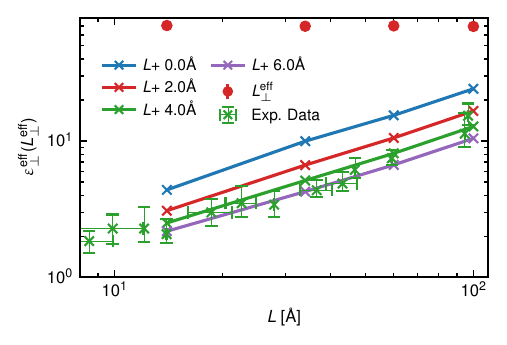}
    \caption{Effective dielectric constant $\epsilon_{\perp} ^{\text{eff}}$ for different possible values of $L_\perp^\text{eff}$ vs.\ the graphene distance $L$.
    The green symbols denote the experimental data by \citeauthor{fumagalli18a}~\cite{fumagalli18a}.
    A best fit to the experimental data is obtained in \cref{fig:capacitance_models} yielding $L_\perp^\text{eff} = L + \SI{3.8}{\angstrom}$.
    }
    \label{fig:effective_medium_equivalent}
\end{figure}

Importantly, in a nanoconfined system the definition of a pore width $h$ is somewhat arbitrary, and typically the distance $L$ between the innermost atoms of the confining walls is used as a measure of pore size---however, this is not always easy to control or measure in experiments.
In theoretical considerations and simulations, there is an additional ambiguity since this typically also is considered to be the distance between the positions where a potential is measured. 
In the experiments, there is a similar issue: there is a distance $h$ measured between the surfaces, whereas the potential is applied between the bottom graphite layer and the AFM tip.
The relation between tip-substrate capacitance $\mathrm{d}C/\mathrm{d}z$ in perpendicular $z$-direction and the effective dielectric constant of a layer of width $h$ is calculated numerically by assuming a model for the geometry and dielectric properties of the tip, the top hBN layer, and their contact \cite{fumagalli18a}.

\Cref{fig:effective_medium_equivalent} compares the experimentally measured effective perpendicular dielectric constant $\epsilon_{\perp}^\text{eff}$ as a function of the graphene distance $L$ (green stars) to our effective dielectric medium following from \cref{eq:perpendicular_macroscopic_profile} (red circles).
The clear mismatch we find can be attributed to several possible origins.
First of all, the distance over which the potential drops could be different, however, as we discuss below, in order to rationalize the experimentally observed reduction of $\varepsilon_\perp^\text{eff}$, this would require a distance larger than $L$ and thus is not in line with the expectations commonly discussed in terms of the Jellium edge \cite{lang73a}.
Second, $h$ could be subject to experimental uncertainties.
Third, there could be an influence due to adsorbed gas or other molecules in the experimental system.
Finally, the mismatch can be due to the assumptions made for the geometry and dielectric properties of the AFM tip and hBN layer, as we further elaborate in the next secion.

In our analysis, such effects reduce to shifts in the stern layer thickness $2\delta_\perp^\mathrm{S} = L-L_\perp^\text{eff}$ that enters the dielectric box model in \cref{eq:perpendicular_macroscopic_profile}.
Whereas this quantity is not known apriori and needs to be determined independently from simulations or experiments, it is imporant to stress that this is different from modifying $L$.
It is thus tempting to study the influence of the choice of the effective length on the effective dielectric constant $\varepsilon_\perp^\text{eff}$.
As seen in \cref{fig:effective_medium_equivalent}, identifying $L_\perp^\text{eff} = L$, leads to a significant and long-range reduction of the effective dielectric constant of the confined water, qualitatively following the experimental observations by \citeauthor{fumagalli18a}~\cite{fumagalli18a}.
Indeed, $L _{\perp}^\text{eff}$ can further be tuned to match these data for $\epsilon_{\perp}^{\text{eff}}$, incorporating the dielectric behavior of the entire system into the value for $\epsilon_{\perp}^\text{eff}$.
This method has been used (with slight modifications) in order to obtain a good fit to the experimental data, i.a.\ by \citeauthor{cox22a}~\cite{cox22a}, \citeauthor{loche20a}~\cite{loche20a} and \citeauthor{becker24a}~\cite{becker24a}, who tuned the effective length of the water slab until it matched the experimental results.
Assuming that the dielectric dividing interface is displaced by $\SI{3}{\angstrom}$ from the graphene sheets away from the water, the experimentally observed dielectric behavior can be reproduced.

However, such a procedure remains empirical and requires experimental data to fit the effective length.
More problematic, relying on this approach to obtain $L_\perp^\text{eff}$ suggests that dielectric constant of confined water is significantly reduced with respect to bulk on separations up to about 100~nm~\cite{fumagalli18a}.
Note again, that here we only played with the position where the elecrostatic potential drop is measured, the dielectric profiles only about a nanometer away from the interface reveal bulk-like behavior (\cref{fig:density_plot}).
We address this apparent contradiction in the following section by explicitly taking into account the Stern layer.

\subsection{Capacitor model and apparent dielectric response}
\begin{figure*}[htbp]
    \centering
    \begin{minipage}[c]{0.49\linewidth}
        \vspace{0pt} 
        \includegraphics[width=\linewidth]{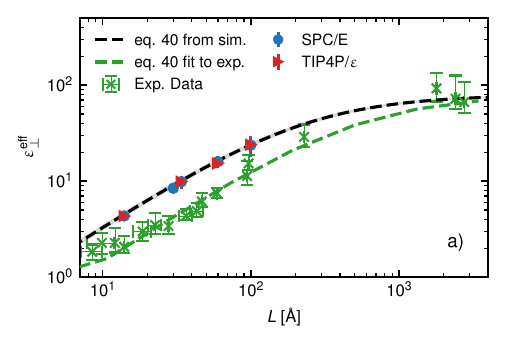}
    \end{minipage}
    \hfill
    \begin{minipage}[c]{0.49\linewidth}
        \vspace{0pt} 
    \def\svgwidth{\columnwidth}
    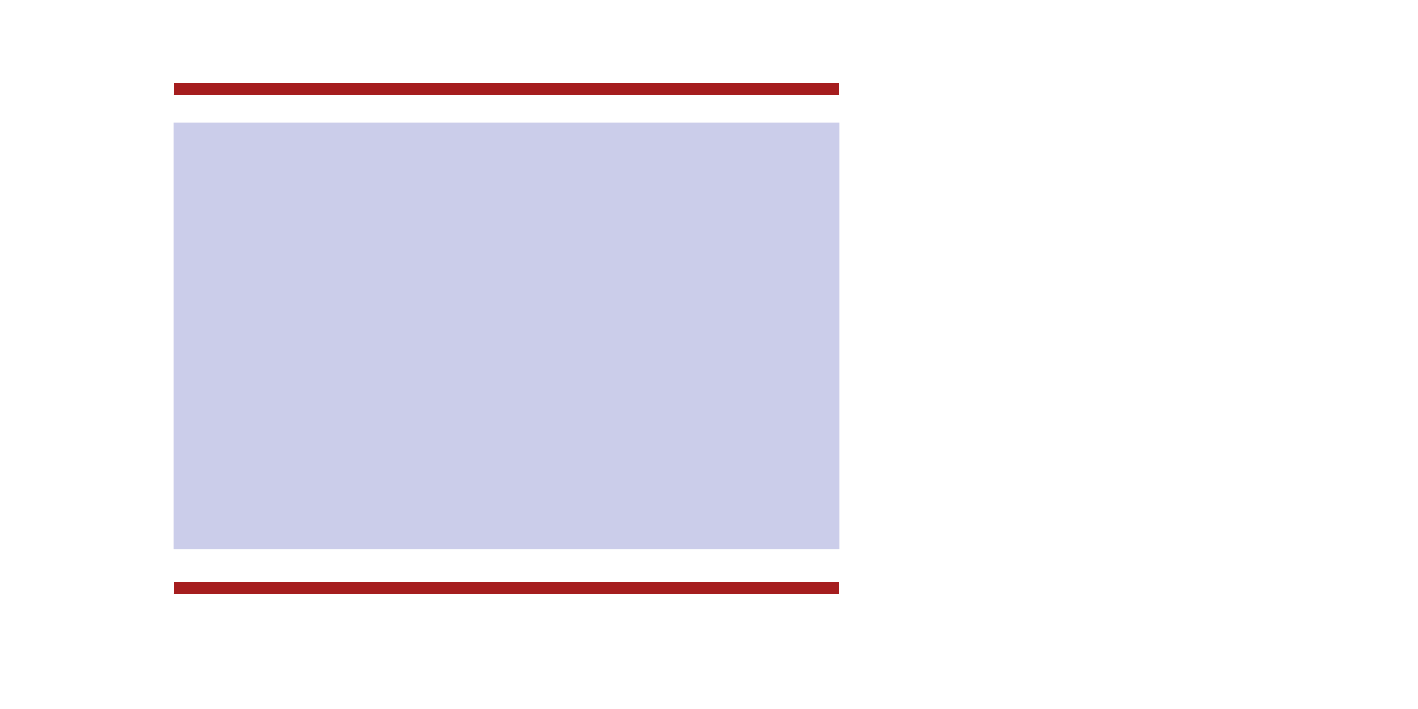

    \end{minipage}
    \caption{
        Panel a): Effective perpendicular dielectric constant obtained from the equivalent circuit model \cref{eq:epsilon_effective_capacitance} (dashed black line) and 
        from our simulations employing different water models (blue and red symbols).
        Green stars denote the results by \citeauthor{fumagalli18a}~\cite{fumagalli18a}.
        The corresponding dashed green line is obtained by including an additional capacitance $C_\mathrm{tip/sub}$---equivalent to increasing the Stern layer thickness by about 1.9~\AA{} on each side, cf.\ green line in \cref{fig:effective_medium_equivalent}---to account for the tip and hBN layer properties in the experimental setup.
        Panel b): Visualization of the equivalent circuit consisting of the water effective perpendicular dielectric response and taking into account the vacuum layer at the interface.
    The resulting capacitance consists of the water layer of width $L_{\perp}^{\text{eff}}$ and dielectric constant $\epsilon_{\perp}^{\text{eff}}$ and two vacuum layers of width $\delta_\perp^\mathrm{S}$ and dielectric constant $\epsilon_{\text{vac}} = 1$.}
    \label{fig:capacitance_models}
\end{figure*}

As already discussed above, one typically probes the potential drop over the entire system, i.e.\ the capacitance of a pore is of interest.
At a hydrophobic surface, such as graphene, water is significantly repelled from the interface, i.e.\ there is a dielectrically dead vacuum layer---akin the Stern layer, see e.g.\ \cref{fig:example_epsilon_profs}.
Correspondingly, the capacitance of this layer needs to be taken into account in the system's total dielectric response in terms of a corresponding equivalent circuit as shown in \cref{fig:capacitance_models}(b).

This equivalent circuit directly results in the following expression:
\begin{equation}
    \label{eq:total_capacitance}
    \frac{1}{C_{\text{tot}} } = \frac{2}{C_{\text{vac}}} + \frac{1}{C_{\text{slab}}},
\end{equation}
where $C_{\text{{vac}}} = \epsilon_0 A/\delta_\perp^{S}$ and $C_{\text{slab}} = \epsilon_0 \epsilon_{\perp}^{\text{{eff}}} A/L_{\perp}^{\text{eff}}$.
Since the microscopic details, i.e.\ the Stern layer thickness $\delta_\perp^{\mathrm{S}}$ and the effective dielectric width $L_\perp^{\mathrm{eff}}$ may not be accessible, one might be tempted to express \cref{eq:total_capacitance} in terms of an apparent dielectric constant corresponding to a single plate capacitor of separation $h$ filled with a homogeneous dielectric medium:
\begin{equation}
    \label{eq:epsilon_effective_capacitance}
\epsilon_\perp ^{\text{eff}} (h) = \frac{h}{2 \delta_\perp^\mathrm{S} + (h - 2 \delta_\perp^\mathrm{S}) / \epsilon_{\perp}^{\text{bulk}}},
\end{equation}
where $\epsilon^{\text{bulk}}_{\perp}$ is the bulk value of the water model.
Note that this expression is equivalent to \cref{eq:perpendicular_macroscopic_profile}, but since only the constant Stern layer thickness $\delta_\perp^\mathrm{S}$ enters, allows for a continous prediction of $\varepsilon_\perp^\text{eff}$.

Previous approaches already typically interpret the experimental data of \citeauthor{fumagalli18a} in terms of such an equivalent circuit, considering the interfacial water instead of the vacuum to contribute one \cite{fumagalli18a, vargheseEffectHydrogenBonds2019} or two capacitors connected in series~\cite{dufils24a, tangScaledependentAnomalousBehavior2024}, whereas the idea of a vacuum layer in connection with a homogeneous dielectric slab has also been proposed based on a mean-field approach \cite{monet21a} or recently in the context of the water's perpendicular averaged dipole density \cite{zubeltzuRedefiningDielectricResponse2025}.

Our approach here is different, since it rigorously eliminates all free parameters, most importantly the identification of the electrode's position in a continuum description of a plate capacitor with the (rather arbitrary) choice of the graphene atoms positions that all of these approaches have in common. 
This stands in contrast to most previous works, where the width (and, in some cases, the dielectric constant) of the interfacial layer is typically fitted in order to agree with measurements.
For example, \citeauthor{fumagalli18a} have proposed a similar model, where an interfacial layer of dielectric constant 2.1 is assumed.
The width of this layer is then fitted to the experimental data, yielding a value of about 7.1~\AA{}.
We also note, that in the experimental setup, only one wall was comprised of graphene, whereas the other wall was made from hBN.

Our simulation data and the prediction resulting from \cref{eq:epsilon_effective_capacitance} are shown in \cref{fig:capacitance_models} and contrasted to the experimental data.
Noteworthy, $\varepsilon_\perp^\text{eff}$ shows a long-ranged and significant reduction of the apparent dielectric constant as the graphene distance $L$ decreases, both in the experimental as well as in our simulation data.
Importantly, our simulation results are independent of the water model used and agree qualitatively excellent with the experimental data.
Possible origins for the remaining quantitative differences are discussed above.
In fact, as shown in \cref{fig:capacitance_models}(a) and in line with the results in \cref{fig:effective_medium_equivalent}, the experimental data can be matched quantitatively if the Stern layer thickness is increased by about $\Delta =1.9$~\AA{} compared to our value reported above, $\delta_\perp^\mathrm{S} = 1.5$~\AA{}.
Or, in other words, an additional capacitances $C_\mathrm{tip/sub}$ that capture all effects due to the tip and substrate dielectric properties and geometry would need to be added in series to \cref{fig:capacitance_models}(b) to match the experimental data quantitatively.

We show in \cref{fig:capacitance_models}(a) the corresponding prediction analogous to \cref{eq:total_capacitance,eq:epsilon_effective_capacitance} which includes the tip/substrate areal capacitance $C_\mathrm{tip/sub}/A = 2\Delta / \varepsilon_0 = 4.66 \mu\,\mathrm{F}/\mathrm{cm}^2$ (dashed green line) in series on both sides of \cref{fig:capacitance_models}(b).
This equivalent circuit corresponds to a total of four capacitors in series: once the effective medium, two times the Stern layer, and once the unknown capacitance anywhere between the AFM tip and the graphite layer.
Summarizing the comparison of our simulations and analysis with experiments, we find striking agreement within the bulk-like effective medium approach if the position at which the potential drop is measured is assumed to be about a molecular layer ($\SI{2}{\angstrom}$) displaced from the positions of the outermost carbon layers.
This in fact reveals striking agreement and can further be interpreted in terms of $C_\mathrm{tip/sub}$.
The simulation results are very accurately reproduced by different water models, indicating universal dielectric behavior.
Further limitations might appear due to the fact that electron spillover is not accounted for in our classical force-field simulations of the electrically inert graphene, but one would actually expect an opposite shift in $\delta_\perp^\mathrm{S}$\cite{lang73a}.
The effect of including the quantum or self-capacitance of the electrode in the equivalent circuit will be addressed in a separate work, yet the presented agreement strongly supports our approach without adjustable parameters.

\section{Summary and Conclusions}
\label{sec:summary}
While the dielectric response of fluids confined in nanometer sized pores is important for a vast amount of processes and technologies, the interpretation and derivation of the dielectric behavior still seems controversial in the literature.
In this work, we systematically revisit the fundamentals starting with the fluid's non-local dielectric response and the derivation of location-dependent, anisotropic dielectric profiles in planar confinement.
Importantly, we discussed the influence of boundary conditions and the corresponding assumptions in detail and showed that the correct treatment of these conditions is important and not trivial.
Furthermore, we emphasize that care has to be taken in the treatment of systems with free charges, which is often neglected~\cite{hu21a,renou15a,jalali21a,zhu22a}.

We presented a systematic approach for deriving macroscopic coarse-grained models from microscopic dielectric profiles.
Applying this method to water confined between graphene sheets to simulations, using the previously untested, but more accurate TIP4P/$\epsilon$ water model, we find that the experimentally observed behavior can be reproduced independently of the water model.
We also find that the problematic issue of finding the correct water number in the simulation of such confined systems can be accounted for by relating its dielectric properties to the corresponding bulk density.

Finally, we reinterpreted the experimentally observed reduction of the dielectric response of confined water in terms of an equivalent circuit that takes into account the dielectrically dead Stern layer at the hydrophobic interface.
This last step puts the experimentally observed dielectric behavior, measured via the capacitance, in line with the microscopic dielectric profiles, which show bulk-like behavior already about $1$~nm away from the interface.

Since determining the dielectric response of confined fluids from molecular simulations is involved and error-prone, we offer the open-source and freely available software library MAICoS~\footnote{\url{https://maicos-analysis.org}} that is intended to provide tools and extended examples for calculating dielectric profiles for planar, cylindrical and spherical symmetries in a repeatable and reliable manner following the FAIR\ principles \cite{wilkinsonFAIRGuidingPrinciples2016}. MAICoS will be detailed in a separate publication.
Thus, we anticipate that this review together with the availability of an open and well-tested software package will help not only to settle differences and controversies in the community, but also to establish a common ground for future studies of the dielectric response of confined fluids.


\section{Data Availability Statement}
The simulation methods for all simulations used for this publication are described in detail in the supporting information~\cite{suppa} and simulation input files that support these findings are openly available at \url{https://doi.org/10.18419/darus-4317}.
All analysis methods with detailed examples employed in this work are available in the software package MAICoS, which is available for download and described under~\url{https://maicos-analysis.org}.

\section{Acknowledgements}
We thank the Deutsche Forschungsgemeinschaft (DFG, German Research Foundation) for funding through Germany's Excellence Strategy - EXC 2075 – 390740016 and SFB 1333/2 – 358283783.
We acknowledge the support by the Stuttgart Center for Simulation Science (SimTech) and state of Baden-Württemberg through bwHPC.
We also acknowledge the scientific exchange and support of the Centre for Molecular Water Science CMWS and the Cluster of Excellence EXC 3120/1 – 533771286.

\section*{Supplementary Material}
In the supplementary material we provide details on the simulation settings and atomistic structures, additional figures, a derivation of the three dimensional fluctuation dissipation equation and details on how the statistical error of the effective dielectric constants were calculated.

%% file: figures/kirk_f_claus.pdf_tex
\begingroup%
  \makeatletter%
  \providecommand\color[2][]{%
    \errmessage{(Inkscape) Color is used for the text in Inkscape, but the package 'color.sty' is not loaded}%
    \renewcommand\color[2][]{}%
  }%
  \providecommand\transparent[1]{%
    \errmessage{(Inkscape) Transparency is used (non-zero) for the text in Inkscape, but the package 'transparent.sty' is not loaded}%
    \renewcommand\transparent[1]{}%
  }%
  \providecommand\rotatebox[2]{#2}%
  \newcommand*\fsize{\dimexpr\f@size pt\relax}%
  \newcommand*\lineheight[1]{\fontsize{\fsize}{#1\fsize}\selectfont}%
  \ifx\svgwidth\undefined%
    \setlength{\unitlength}{732.15001384bp}%
    \ifx\svgscale\undefined%
      \relax%
    \else%
      \setlength{\unitlength}{\unitlength * \real{\svgscale}}%
    \fi%
  \else%
    \setlength{\unitlength}{\svgwidth}%
  \fi%
  \global\let\svgwidth\undefined%
  \global\let\svgscale\undefined%
  \makeatother%
  \begin{picture}(1,0.39052574)%
    \lineheight{1}%
    \setlength\tabcolsep{0pt}%
    \put(0,0){\includegraphics[width=\unitlength,page=1]{kirk_f_claus.pdf}}%
    \put(0.35370108,0.27637025){\color[rgb]{0.7372549,0.12941176,0.13333333}\makebox(0,0)[lt]{\lineheight{1.25}\smash{\begin{tabular}[t]{l}$\epsilon$\end{tabular}}}}%
    \put(0.29916038,0.07429319){\color[rgb]{0.7372549,0.12941176,0.13333333}\makebox(0,0)[lt]{\lineheight{1.25}\smash{\begin{tabular}[t]{l}$\epsilon_{\mathrm{RF}} \approx \epsilon$\end{tabular}}}}%
    \put(0,0){\includegraphics[width=\unitlength,page=2]{kirk_f_claus.pdf}}%
    \put(0.05822939,0.28980874){\color[rgb]{0.7372549,0.12941176,0.13333333}\makebox(0,0)[lt]{\lineheight{1.25}\smash{\begin{tabular}[t]{l}$\epsilon$\end{tabular}}}}%
    \put(0.02340436,0.07500402){\color[rgb]{0.7372549,0.12941176,0.13333333}\makebox(0,0)[lt]{\lineheight{1.25}\smash{\begin{tabular}[t]{l}$\epsilon_{\mathrm{RF}} = 1$\end{tabular}}}}%
    \put(-0.00110029,0.32112666){\color[rgb]{0,0,0}\makebox(0,0)[lt]{\lineheight{1.25}\smash{\begin{tabular}[t]{l}{\normalsize a)}\end{tabular}}}}%
    \put(0.25346526,0.32388532){\color[rgb]{0,0,0}\makebox(0,0)[lt]{\lineheight{1.25}\smash{\begin{tabular}[t]{l}{\normalsize b)}\end{tabular}}}}%
    \put(0,0){\includegraphics[width=\unitlength,page=3]{kirk_f_claus.pdf}}%
    \put(0.59700877,0.33651468){\color[rgb]{0,0,0}\makebox(0,0)[lt]{\lineheight{1.25}\smash{\begin{tabular}[t]{l}{\normalsize c)}\end{tabular}}}}%
    \put(0.66971619,0.05643022){\color[rgb]{0.7372549,0.12941176,0.13333333}\makebox(0,0)[lt]{\lineheight{1.25}\smash{\begin{tabular}[t]{l}$\epsilon$\end{tabular}}}}%
    \put(0.91610835,0.05206877){\color[rgb]{0.7372549,0.12941176,0.13333333}\makebox(0,0)[lt]{\lineheight{1.25}\smash{\begin{tabular}[t]{l}$\epsilon_{\mathrm{RF}}$\end{tabular}}}}%
  \end{picture}%
\endgroup%

%% file: figures/virtual_cutting.pdf_tex
\begingroup%
  \makeatletter%
  \providecommand\color[2][]{%
    \errmessage{(Inkscape) Color is used for the text in Inkscape, but the package 'color.sty' is not loaded}%
    \renewcommand\color[2][]{}%
  }%
  \providecommand\transparent[1]{%
    \errmessage{(Inkscape) Transparency is used (non-zero) for the text in Inkscape, but the package 'transparent.sty' is not loaded}%
    \renewcommand\transparent[1]{}%
  }%
  \providecommand\rotatebox[2]{#2}%
  \newcommand*\fsize{\dimexpr\f@size pt\relax}%
  \newcommand*\lineheight[1]{\fontsize{\fsize}{#1\fsize}\selectfont}%
  \ifx\svgwidth\undefined%
    \setlength{\unitlength}{655.5187748bp}%
    \ifx\svgscale\undefined%
      \relax%
    \else%
      \setlength{\unitlength}{\unitlength * \real{\svgscale}}%
    \fi%
  \else%
    \setlength{\unitlength}{\svgwidth}%
  \fi%
  \global\let\svgwidth\undefined%
  \global\let\svgscale\undefined%
  \makeatother%
  \begin{picture}(1,0.471798)%
    \lineheight{1}%
    \setlength\tabcolsep{0pt}%
    \put(0,0){\includegraphics[width=\unitlength,page=1]{virtual_cutting.pdf}}%
    \put(0.61507748,0.43531946){\color[rgb]{0,0,0}\makebox(0,0)[lt]{\lineheight{1.25}\smash{\begin{tabular}[t]{l}$z$\end{tabular}}}}%
    \put(0,0){\includegraphics[width=\unitlength,page=2]{virtual_cutting.pdf}}%
    \put(0.8708092,0.41259346){\color[rgb]{0,0,0}\makebox(0,0)[lt]{\lineheight{1.25}\smash{\begin{tabular}[t]{l}$x,y$\end{tabular}}}}%
    \put(0.41257545,0.3957117){\color[rgb]{0,0,0}\makebox(0,0)[lt]{\lineheight{1.25}\smash{\begin{tabular}[t]{l}$\rho_{\mathrm{cut}}(a, z)$\end{tabular}}}}%
    \put(0.54461218,0.01006483){\color[rgb]{0.24313725,0.50980392,0.0745098}\makebox(0,0)[lt]{\lineheight{1.25}\smash{\begin{tabular}[t]{l}$a_{\text{cut}}$\end{tabular}}}}%
  \end{picture}%
\endgroup%

%% file: figures/effective_medium_equivalent.pdf_tex
\begingroup%
  \makeatletter%
  \providecommand\color[2][]{%
    \errmessage{(Inkscape) Color is used for the text in Inkscape, but the package 'color.sty' is not loaded}%
    \renewcommand\color[2][]{}%
  }%
  \providecommand\transparent[1]{%
    \errmessage{(Inkscape) Transparency is used (non-zero) for the text in Inkscape, but the package 'transparent.sty' is not loaded}%
    \renewcommand\transparent[1]{}%
  }%
  \providecommand\rotatebox[2]{#2}%
  \newcommand*\fsize{\dimexpr\f@size pt\relax}%
  \newcommand*\lineheight[1]{\fontsize{\fsize}{#1\fsize}\selectfont}%
  \ifx\svgwidth\undefined%
    \setlength{\unitlength}{680.31496063bp}%
    \ifx\svgscale\undefined%
      \relax%
    \else%
      \setlength{\unitlength}{\unitlength * \real{\svgscale}}%
    \fi%
  \else%
    \setlength{\unitlength}{\svgwidth}%
  \fi%
  \global\let\svgwidth\undefined%
  \global\let\svgscale\undefined%
  \makeatother%
  \begin{picture}(1,0.5)%
    \lineheight{1}%
    \setlength\tabcolsep{0pt}%
    \put(0,0){\includegraphics[width=\unitlength,page=1]{effective_medium_equivalent.pdf}}%
    \put(0.30453995,0.25610229){\color[rgb]{0.14509804,0.16862745,0.50980392}\makebox(0,0)[lt]{\lineheight{1.25}\smash{\begin{tabular}[t]{l}$\epsilon_{\perp}^{\mathrm{eff}}$\end{tabular}}}}%
    \put(0,0){\includegraphics[width=\unitlength,page=2]{effective_medium_equivalent.pdf}}%
    \put(0.64467426,0.24670706){\color[rgb]{0,0,0}\rotatebox{90}{\makebox(0,0)[lt]{\lineheight{1.25}\smash{\begin{tabular}[t]{l}$L$\end{tabular}}}}}%
    \put(0.04999188,0.4201585){\color[rgb]{0.7372549,0.12941176,0.13333333}\makebox(0,0)[lt]{\lineheight{1.25}\smash{\begin{tabular}[t]{l}$\delta_\perp ^{\text{s}}$\end{tabular}}}}%
    \put(0.04999188,0.09037907){\color[rgb]{0.7372549,0.12941176,0.13333333}\makebox(0,0)[lt]{\lineheight{1.25}\smash{\begin{tabular}[t]{l}$\delta_\perp ^{\text{s}}$\end{tabular}}}}%
    \put(0.01474318,0.47435682){\color[rgb]{0,0,0}\makebox(0,0)[lt]{\lineheight{1.25}\smash{\begin{tabular}[t]{l}b)\end{tabular}}}}%
    \put(0,0){\includegraphics[width=\unitlength,page=3]{effective_medium_equivalent.pdf}}%
    \put(0.00879898,0.23976146){\color[rgb]{0.55294118,0.56862745,0.81568627}\makebox(0,0)[lt]{\lineheight{1.25}\smash{\begin{tabular}[t]{l}$L_{\perp}^{\mathrm{eff}}$\end{tabular}}}}%
    \put(0,0){\includegraphics[width=\unitlength,page=4]{effective_medium_equivalent.pdf}}%
    \put(0.82261918,0.42461987){\color[rgb]{0.64705882,0.11372549,0.11764706}\makebox(0,0)[lt]{\lineheight{1.25}\smash{\begin{tabular}[t]{l}$C_{\mathrm{vac}}$\end{tabular}}}}%
    \put(0.82151051,0.25333068){\color[rgb]{0.55294118,0.56862745,0.81568627}\makebox(0,0)[lt]{\lineheight{1.25}\smash{\begin{tabular}[t]{l}$C_{\perp}^{\mathrm{eff}}$\end{tabular}}}}%
    \put(0.81929315,0.08204142){\color[rgb]{0.64705882,0.11372549,0.11764706}\makebox(0,0)[lt]{\lineheight{1.25}\smash{\begin{tabular}[t]{l}$C_{\mathrm{vac}}$\end{tabular}}}}%
  \end{picture}%
\endgroup%

%% file: figures/piston_setup.pdf_tex
\begingroup%
  \makeatletter%
  \providecommand\color[2][]{%
    \errmessage{(Inkscape) Color is used for the text in Inkscape, but the package 'color.sty' is not loaded}%
    \renewcommand\color[2][]{}%
  }%
  \providecommand\transparent[1]{%
    \errmessage{(Inkscape) Transparency is used (non-zero) for the text in Inkscape, but the package 'transparent.sty' is not loaded}%
    \renewcommand\transparent[1]{}%
  }%
  \providecommand\rotatebox[2]{#2}%
  \newcommand*\fsize{\dimexpr\f@size pt\relax}%
  \newcommand*\lineheight[1]{\fontsize{\fsize}{#1\fsize}\selectfont}%
  \ifx\svgwidth\undefined%
    \setlength{\unitlength}{1585.77525221bp}%
    \ifx\svgscale\undefined%
      \relax%
    \else%
      \setlength{\unitlength}{\unitlength * \real{\svgscale}}%
    \fi%
  \else%
    \setlength{\unitlength}{\svgwidth}%
  \fi%
  \global\let\svgwidth\undefined%
  \global\let\svgscale\undefined%
  \makeatother%
  \begin{picture}(1,0.4540366)%
    \lineheight{1}%
    \setlength\tabcolsep{0pt}%
    \put(0,0){\includegraphics[width=\unitlength,page=1]{piston_setup.pdf}}%
    \put(0.30369532,0.05359683){\color[rgb]{0.22745098,0.48627451,0.05490196}\makebox(0,0)[lt]{\lineheight{1.25}\smash{\begin{tabular}[t]{l}$H _{\text{channel}}$\end{tabular}}}}%
    \put(0,0){\includegraphics[width=\unitlength,page=2]{piston_setup.pdf}}%
    \put(-0.00156653,0.23884131){\color[rgb]{0.7372549,0.12941176,0.13333333}\makebox(0,0)[lt]{\lineheight{1.25}\smash{\begin{tabular}[t]{l}$L$\end{tabular}}}}%
    \put(0,0){\includegraphics[width=\unitlength,page=3]{piston_setup.pdf}}%
    \put(0.88788393,0.22459943){\color[rgb]{0,0.05098039,0.36470588}\makebox(0,0)[lt]{\lineheight{1.25}\smash{\begin{tabular}[t]{l}$P = \SI{1}{\bar}$\end{tabular}}}}%
    \put(0.25967512,0.4028784){\color[rgb]{0,0,0}\makebox(0,0)[lt]{\lineheight{1.25}\smash{\begin{tabular}[t]{l}$H _{\text{slab}} = \SI{50}{\angstrom}$\end{tabular}}}}%
  \end{picture}%
\endgroup%